\documentclass[journal,10pt,]{IEEEtran}
\IEEEoverridecommandlockouts

\usepackage{algorithmic}
\usepackage{textcomp}

\usepackage{graphicx,subcaption}
\usepackage{amsmath,amsthm,amsfonts,amssymb}
\usepackage{cite,soul}
\usepackage{bm}
\usepackage{bbm}
\usepackage{url}
\usepackage{array}
\usepackage{color}
\usepackage{multirow}
\usepackage{booktabs}
\usepackage[table,xcdraw]{xcolor}
\usepackage{enumitem}

\usepackage{makecell}

\theoremstyle{plain}
\newtheorem{thm}{Theorem}
\newtheorem{lem}[thm]{Lemma}

\newtheorem{prop}[thm]{Proposition}
\newtheorem{cor}[thm]{Corollary}

\newtheorem{sty1}{Theorem}
\newtheorem{defi}[sty1]{Definition}

\newenvironment{NewProof}{{\noindent\it Proof.}}{\hfill $\blacksquare$\par}

\begin{document}

\title{Channel Cycle Time: A New Measure of Short-term Fairness}

\author{%

   \IEEEauthorblockN{Pengfei Shen, Yulin Shao, Haoyuan Pan, Lu Lu, Yonina C. Eldar}

   
   

   
\thanks{ 
P. Shen and L. Lu are with the University of Chinese Academy of Sciences, and the Key Laboratory of Space Utilization, Chinese Academy of Sciences, Beijing 100094, China (emails: \{shenpengfei19, lulu\}@csu.ac.cn).

Y. Shao is with the State Key Laboratory of Internet of Things for Smart City, University of Macau, Macau S.A.R, China. He is also with the Department of Electrical and Electronic Engineering, Imperial College London, London SW7 2AZ, U.K. (e-mail: ylshao@ieee.org).

H. Pan is with the College of Computer Science and Software Engineering, Shenzhen University, Shenzhen, China (e-mail: hypan@szu.edu.cn). 

Y. Eldar is with the Weizmann Institute of Science, Rehovot 7610001, Israel (e-mail:yonina.eldar@weizmann.ac.il).
}
}

\maketitle

\begin{abstract}
This paper puts forth a new metric, dubbed channel cycle time (CCT), to measure the short-term fairness of communication networks. CCT characterizes the average duration between two consecutive successful transmissions of a user, during which all other users successfully accessed the channel at least once. In contrast to existing short-term fairness measures, CCT provides more comprehensive insight into the transient dynamics of communication networks, with a particular focus on users' delays and jitter. To validate the efficacy of our approach, we analytically characterize the CCTs for two classical communication protocols: slotted Aloha and CSMA/CA. The analysis demonstrates that CSMA/CA exhibits superior short-term fairness over slotted Aloha. Beyond its role as a measurement metric, CCT has broader implications as a guiding principle for the design of future communication networks by emphasizing factors like fairness, delay, and jitter in short-term behaviors.
\end{abstract}


\section{Introduction}\label{sec:I}
Fairness refers to the principle of providing equal access to resources without discrimination \cite{5461911,6517050,8665952}. In multiple-access networks, where multiple users share the scarce radio frequency resource and simultaneous transmissions result in collisions, fairness is a critical design principle that guarantees the timely and reliable communication of each user, without suffering undue delay or interference from other users \cite{koksal2000analysis,1378897,5062017,8902705,shao2021federated}.

Fairness in network resource allocation can be classified as long-term or short-term, depending on the time scale over which channel resources are allocated \cite{6778811,5062022,6816520,shao2020flexible}.
\begin{itemize}[leftmargin=0.45cm]
    \item Long-term fairness is concerned with allocating channel resources fairly over a sufficiently long period of time. It characterizes the ergodic behavior of a multiple-access control (MAC) protocol and is the predominant focus of the current literature.
    \item Short-term fairness, in contrast, is concerned with ensuring the fair allocation of network resources in the immediate or near term. It characterizes the transient behavior of a MAC protocol. 
\end{itemize}

Achieving short-term fairness is more challenging. A short-term fair MAC protocol is also long-term fair, but the reverse is not always true. A user may monopolize the channel in short periods to the detriment of other users, even if it has the same channel occupancy rate as other users over a long period.

The proliferation of machine-type and low-latency applications has catalyzed a notable shift in the evaluation of MAC protocols, with an increasing emphasis on short-term fairness over long-term fairness. This shift recognizes the need for MAC protocols to prioritize near-term performance and responsiveness in future networks, where delays and latency can have significant impact on user experience.

To measure short-term fairness, \cite{1378897,berger2005short} proposed a fairness index: the number of inter-transmissions that other hosts may perform between two consecutive transmissions of a given host. 
Consider a multiple-access network with three users A, B, and C, and a pattern of successful transmissions `ABCABBCBAC'.
The number of inter-transmissions for users A, B, and C is \{2, 4\}, \{2, 0, 1\}, and \{3, 2\}, respectively. 
The authors proposed to characterize the short-term fairness by the probability density function (PDF) of the number of inter-transmissions of all users.
This approach has two main limitations.
First, it considers only successful transmissions, ignoring the time consumed by these transmissions. As a consequence, a poorly designed MAC protocol with frequent collisions could have the same short-term fairness as a well-designed MAC protocol, if measured by the number of inter-transmissions. 
Second, obtaining the entire PDF of the number of inter-transmissions can be cumbersome, especially when comparing the short-term fairness of two MAC protocols. In \cite{1378897}, the authors proposed using the average number of inter-transmissions as a simpler way to evaluate short-term fairness, with a smaller value indicating better fairness. However, the average number of inter-transmissions loses too much information compared to the entire PDF (see the example in Section~\ref{sec:II-B}), hence is inadequate to fully characterize the short-term fairness of a MAC protocol.

The authors in \cite{koksal2000analysis} proposed a ``renewal reward method'' to measure short-term fairness, where the transitions of channel occupation are assigned rewards.
In particular, the reward a user receives from a successful transmission depends on the number of transmissions of other users from the user’s last successful transmission. 
The reward function is designed to be non-decreasing and reaches the minimum when the same user keeps the channel.
A MAC protocol that yields higher rewards is considered short-term fairer.
However, similar to the number of inter-transmissions, the renewal reward method considers only successful transmissions. It further requires the reward function to be carefully designed for individual MAC protocols to be evaluated.

Another, and perhaps the simplest approach to measure short-term fairness is to reuse the long-term fairness measures in a short period of time. Specifically, for each epoch, we count the number of successful transmissions of each user over a short period and compute the $\alpha$ fairness \cite{5062017,shao2022learning}, Jain's index \cite{jain1984quantitative,8746330}, or any other traditional fairness measure \cite{5461911,4346554,shao2020significant,kelly1997charging,9356328,li2019adaptive,d2020fairness} to capture the short-term behaviors of the network. 
Compared with aforementioned approaches that consider only successful transmissions, this method captures the transient behavior of the network by setting a time window to count the number of successful transmissions.
The challenge, however, lies in determining an appropriate duration of the time window.
At different epochs, the network state varies, hence the amount of time required to measure the transient behavior of the network is different. For a given MAC protocol, determining the appropriate time window to evaluate short-term fairness can be a non-trivial problem in itself.

In this paper, we put forth a new metric, dubbed channel cycle time (CCT), to measure short-term fairness. CCT is a time measure that characterizes the average duration between two successful transmissions of a user, during which all other users have successfully accessed the channel at least once. Compared with existing short-term fairness measures, our measure has three salient features.
\begin{itemize}[leftmargin=0.45cm]
    \item CCT is a single real value that is easy to compute. This facilitates easy comparison of the short-term performance of different MAC protocols.
    \item CCT effectively captures the transient behavior of a MAC protocol under varying network states, reflecting the average time it takes for all users to successfully transmit at least once.
    \item CCT provides a comprehensive picture of the short-term fairness of a MAC protocol, with an emphasis on users' delay and jitter. 
\end{itemize}

To demonstrate the effectiveness of our new approach, we consider two homogeneous multiple-access networks operated with two classical MAC protocols: slotted Aloha and carrier-sense multiple access with collision avoidance (CSMA/CA), respectively.
The closed-form CCT is derived for both cases.
It is shown that CSMA/CA is a short-term fairer protocol than slotted Aloha.

Beyond its role as a short-term fairness measurement, CCT can provide guidelines for MAC protocol development.
For both homogeneous networks employing slotted Aloha and CSMA/CA, we optimize channel access parameters, such as transmission probability and contention window size, using the CCT, yielding revamped versions of slotted Aloha and CSMA/CA that place a premium on short-term fairness.

Further, we investigate a two-user heterogeneous network, wherein one user employs CSMA/CA and the other, an intelligent user with adaptive learning capabilities, adjusts its behavior in response. 
By setting the CCT as the optimization goal, we analytically derive the minimum achievable CCT in such a two-user network.
When juxtaposed against the homogeneous network where both users employ CSMA/CA, the heterogeneous configuration remarkably reduces the CCT by $34.57\%$. 
This reduction underscores the potential of CCT as a transformative force in shaping short-term fairness in future multi-user networks.

The remainder of this paper is structured as follows:
Section~\ref{sec:II} formally defines CCT and reveals its superiority over existing measures.
Sections~\ref{sec:III} and \ref{sec:IV} delve into the analysis of CCT in two homogeneous networks operating with slotted Aloha and CSMA/CA, respectively.
Section~\ref{sec:V} explores the optimal CCT in a two-user heterogeneous network.
Analytical and simulation results are presented in Section~\ref{sec:VI}.
Section~\ref{sec:VII} concludes this paper.


\section{Channel Cycle Time}\label{sec:II}
\subsection{Definition of CCT}\label{sec:II-A}
This section formally introduces the concept of channel cycle time.
We consider a multiple-access network, wherein $N$ users communicate with a common access point (AP) in a shared channel using a given MAC protocol.
To start with, we define the ``refresh moment'', ``refresh time'', and ``cycle time'' for each user.
\begin{defi}[Refresh moment]\label{defi:refresh_moment}
A moment is a refresh moment of a user if and only if
\begin{itemize}[leftmargin=0.35cm]
    \item A successful channel access of the user ends at this moment.
    \item The next successful channel access belongs to other users.
\end{itemize}
\end{defi}

\begin{defi}[Refresh time]\label{defi:refresh_time}
A refresh time of the $n$-th user, denoted by $\mathcal{T}_n$, is defined as the time between two consecutive refresh moments of the $n$-th user.
\end{defi}

\begin{figure}[t]
\centering
\includegraphics[width=0.9\columnwidth]{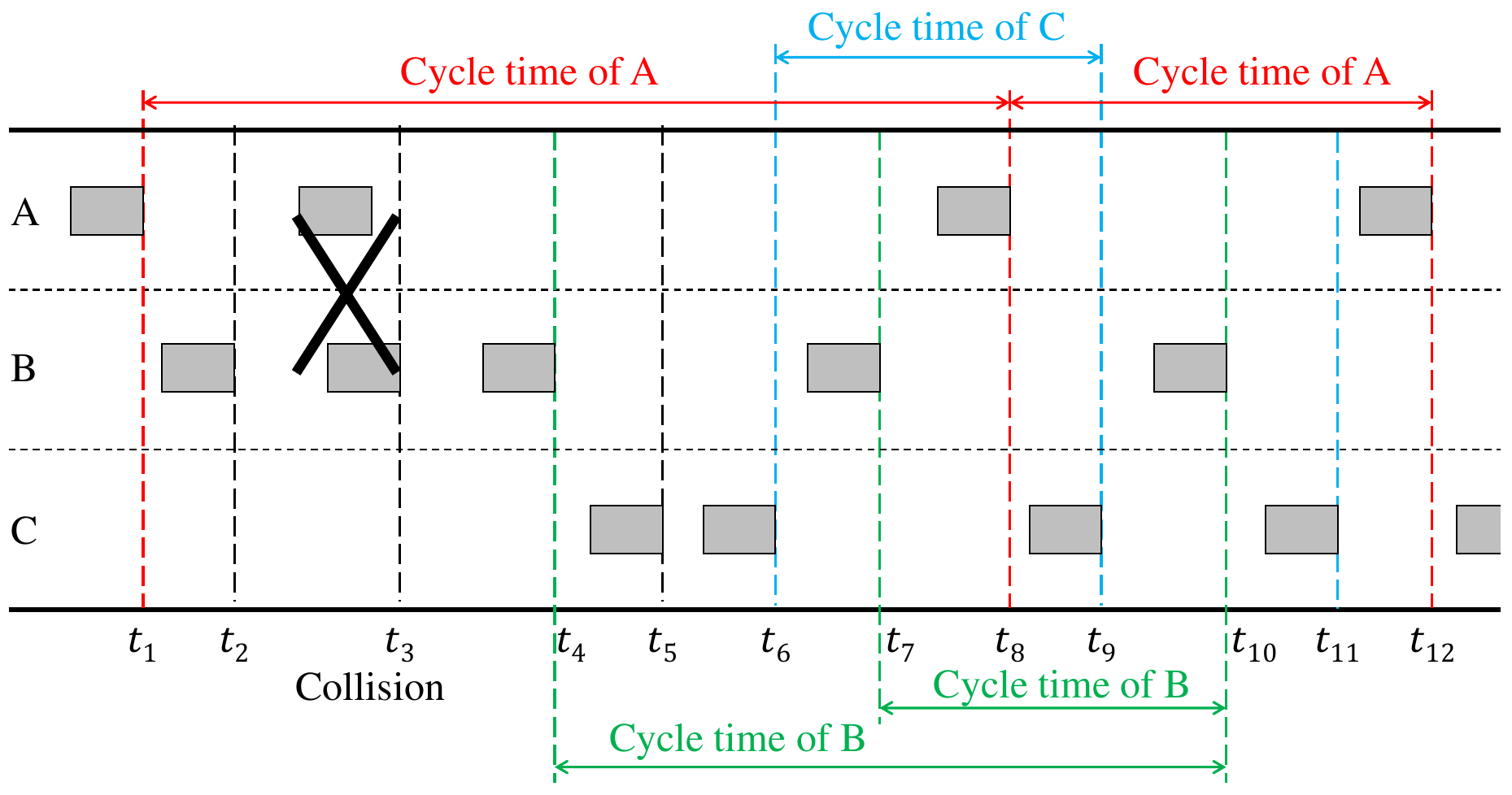}
\caption{Illustrations of the refresh moment and cycle time of users in a multiple-access network ($N=3$).}
\label{fig:refresh_moment}
\end{figure}

An example is given in Fig.~\ref{fig:refresh_moment}, where there are three users A, B, and C in the network.
As can be seen, the pattern of successful transmissions is `ABBCCBACBCA'. Based on Definitions~\ref{defi:refresh_moment} and \ref{defi:refresh_time}, the refresh moments of users A, B, and C are $\{t_1, t_8, t_{12}\}$, $\{t_4, t_7, t_{10}\}$, and $\{t_6, t_9, t_{11}\}$, respectively, and the corresponding refresh times are $\{t_8-t_1, t_{12}-t_8\}$, $\{t_7-t_4, t_{10}-t_7\}$, and $\{t_9-t_6, t_{11}-t_9\}$, respectively.

Consider any two epochs $t$ and $t^\prime$, we denote by $\{M_{n^\prime}(t,t^\prime):n^\prime=1,2,...,N\}$ the number of successful channel accesses of all users between $t$ and $t^\prime$. The cycle time of a user can be defined as follows.

\begin{defi}[Cycle time]\label{defi:cycle_time}
Consider two refresh moments $t_0$ and $t_1$, $t_0<t_1$, of the $n$-th user. The cycle time of the $n$-th user, denoted by $\Gamma_{n}$, is a random variable.
\begin{itemize}[leftmargin=0.45cm]
    \item When $t_0$ and $t_1$ are consecutive refresh moments, $|t_0-t_1|$ is a cycle time of the $n$-th user if and only if
\begin{eqnarray} \label{eq:II-1}
\min_{n^\prime} M_{n^\prime}(t_0,t_1) > 0.
\end{eqnarray}
\item When $t_0$ and $t_1$ are nonconsecutive, $|t_0-t_1|$ is a cycle time of the $n$-th user if and only if
\begin{eqnarray}\label{eq:II-2}
\begin{cases}
\min_{n^\prime} M_{n^\prime}(t_0,t_1) > 0, \\
\min_{n^\prime} M_{n^\prime}(t_0,t^\prime) = 0,
\end{cases}
\end{eqnarray}
where $t': t_0<t'<t_1$ is the closest refresh moment of the $n$-th user to $t_1$.
\end{itemize}
\end{defi}

Succinctly speaking, $\Gamma_{n}$ measures the duration between two closest refresh moments of the $n$-th user, in which all other users successfully access the channel at least once. 
In Fig.~\ref{fig:refresh_moment}, for example, the cycle times of users A, B, and C are $\{t_8-t_1, t_{12}-t_8\}$, $\{t_{10}-t_4,t_{10}-t_7\}$, and $\{t_9-t_6\}$, respectively.
Note that $t_7-t_4$ is not a cycle time of B, because there is no successful transmission from user A in this period;
$t_{11}-t_6$ is also not a cycle time of C, because \eqref{eq:II-2} is not satisfied.

Definition~\ref{defi:cycle_time} also indicates that the cycle time of a user consists of several refresh time. That is,
\begin{equation}\label{eq:III-1}
    \Gamma_n = \sum_{i=1}^{I_r} \mathcal{T}_{n,i},
\end{equation}
where $I_r$ is the number of refresh time that $\Gamma_n$ contains. Therefore, the average cycle time can be calculated as
\begin{equation}\label{eq:III-2}
\mathbb{E}[\Gamma_n]=\mathbb{E}_{I_r, \mathcal{T}_{n,i}}\left[\sum_{i=1}^{I_r} \mathcal{T}_{n,i}\right].
\end{equation}

\begin{defi}[Channel cycle time]\label{defi:channel_cycle_time}
In a multiple-access network with $N$ users, suppose the users go through $\xi_n(T)$, $n=1,2,...,N$, cycle times within a period of time $T$. The channel cycle time (CCT) of the network is defined as the average cycle time of all users:
\begin{eqnarray}\label{eq:II-3}
\Psi = \lim_{T\to\infty} \frac{\sum_{n=1}^{N} \xi_n(T)\mathbb{E}[\Gamma_n]}{\sum_{n=1}^{N}\xi_n(T)}.
\end{eqnarray}
\end{defi}

Just like existing short-term measures, calculating CCT for a general stationary MAC protocol necessitates a historical record, specifically capturing the channel access history of all network users within this timeframe. 
Within this historical record, each user goes through multiple cycle times.
To compute the CCT, we count the number of cycle times experienced by each user $\xi_n$ and calculate the average cycle time $\mathbb{E}[\Gamma_n]$ for each user. Finally, CCT can be obatined following \eqref{eq:II-3}.
For certain MAC protocols, it is possible to derive CCT in closed-form.

\begin{cor}\label{thm:cor1}
In a multiple-access network with two users A and B, we have
\begin{equation} \label{eq:II-4}
\Psi = \mathbb{E}[\Gamma_A]=\mathbb{E}[\Gamma_B].
\end{equation}
\end{cor}

\begin{NewProof}
We first prove that $\mathbb{E}[\Gamma_A]=\mathbb{E}[\Gamma_B]$ in a two user network. Consider a general transmission pattern:
\begin{equation*}
\cdots\!\underbrace{A\cdots A}_{t_{A1}}
\!\underbrace{B\cdots B}_{t_{B1}}
\!\underbrace{A\cdots A}_{t_{A2}}
\!\underbrace{B\cdots B}_{t_{B2}}
\!\underbrace{A\cdots A}_{t_{A3}}
\!\underbrace{B\cdots B}_{t_{B3}}
\cdots,
\end{equation*}
where $t_{Ai}$ and $t_{Bi}$ ($i=1,2,\cdots$) are the time consumed in the corresponding periods.
Then cycle times of user A are
$t_{B1}+t_{A2},t_{B2}+t_{A3},\cdots$, which are also cycle times of user B, and vice versa.
That is, the cycle time values of A and B are the same.
As a result, we have $\mathbb{E}[\Gamma_A]=\mathbb{E}[\Gamma_B]$.
Then, \eqref{eq:II-4} follows from the definition of CCT in \eqref{eq:II-3}.
\end{NewProof}

For any MAC protocol, a smaller CCT indicates better short-term fairness. In particular, when the packet duration of each user is fixed, round-robin TDMA is the short-term fairest protocol.
In this case, CCT is lower bounded by $\Psi \geq \sum_{n=1}^N \ell_{n}$,
where $\ell_n$ denotes the duration of the data packet of the $n$-th user.

\subsection{Superiority of CCT over existing measures}\label{sec:II-B}
Consider a network with two users A and B, the packet duration of which is $\ell_A$ and $\ell_B$, respectively.
Suppose there are two TDMA protocols: the first protocol operates as `AABBAABB...', while the second operates as `ABABABAB...'
Which protocol is fairer?

It is easy to see that the two protocols are equally fair in the long term.
To measure the short-term fairness,
\begin{itemize}[leftmargin=0.45cm]
    \item If we use the average number of inter-transmissions, both protocols have a measure of 1, meaning that they are equally fair.
    \item In contrast, the CCT of the two protocols are $2(\ell_A+\ell_B)$ and $(\ell_A+\ell_B)$, respectively, indicating that the second MAC protocol is short-term fairer, which is more in line with our intuition.
    \item Finally, if we use traditional long-term fairness measures with a time window, the choice of window size has a significant impact on the results obtained, as stated in the introduction. Specifically, if we choose the window size to be $2(\ell_A+\ell_B)$, the two protocols are equally fair. Varying the window size yields different results.
\end{itemize}

In the following two sections, we delve into homogeneous networks operated with two classical MAC protocols: slotted Aloha and CSMA/CA. Our focus will be on analyzing their short-term fairness via CCT. To extract both analytical findings and deeper insights, we will consistently approach a saturated traffic scenario, wherein each user's queue is brimming and there is always a packet ready for transmission.
Furthermore, we assume that the duration of data packets for each user remains consistent and is denoted by $\ell_{\text{pkt}}$. 

As stated in the introduction, CCT can be used as not only an evaluation tool, but also a guiding principle for optimizing classical MAC protocols or crafting novel protocols that prioritize short-term fairness. Subsequent sections will delve deeper into this as well.

\section{CCT of Slotted Aloha}\label{sec:III}
Slotted Aloha is a simple and efficient MAC protocol.
When operated with slotted Aloha, time is divided into equal-sized slots with duration $T_\text{slot}$, and users can transmit only at the beginning of a time slot.
Consider a network with $N$ users and let $T_\text{slot}=\ell_\text{pkt}$.
At the beginning of any time slot, each user transmit a packet with probability $p$.
A packet can be successfully transmitted only when all other users are silent in this slot (the probability of which is $Np(1-p)^{N-1}$).

\subsection{CCT of slotted Aloha}\label{sec:III-A}
In slotted Aloha, the transmissions across different slots are independent.
Thus, in a cycle time, $\mathcal{T}_{n,i}, i=1,2,\cdots,I_r$ are independent and identically distributed (i.i.d.) random variables. Eq. \eqref{eq:III-2} can be refined as
\begin{equation}\label{eq:III-3}
\mathbb{E}[\Gamma_n]=\mathbb{E}[I_r] \cdot \mathbb{E}[\mathcal{T}_n].
\end{equation}
The average cycle time of a user can be obtained by deriving $\mathbb{E}[I_r]$ and $\mathbb{E}[\mathcal{T}_n]$, respectively.
\begin{lem}\label{lem:average_time_period_ALOHA}
In slotted Aloha, the average time it takes to successfully transmit a packet is
\begin{equation}\label{eq:III-4}
\overline{T} = \frac{T_\text{slot}}{Np(1-p)^{N-1}}.
\end{equation}
\end{lem}

\begin{NewProof}
Assume there are $K$ slots between two consecutive successful transmissions. The distribution of $K$ is given by
\begin{equation*}
    \Pr(K\!=\!k)\!=\![1\!-\!Np(1\!-\!p)^{N\!-\!1}]^{k\!-\!1} \!\cdot\! Np(1\!-\!p)^{N\!-\!1}, k\!=\!1,2,\cdots
\end{equation*}
Thus, we have $\overline{T} = \mathbb{E}(K) \cdot T_\text{slot}$, which gives us \eqref{eq:III-4}.
\end{NewProof}

\begin{prop}\label{thm:RRT_ALOHA}
When operated with slotted Aloha, the average refresh time of each user is given by
\begin{equation}\label{eq:III-5}
    \mathbb{E}[\mathcal{T}]=\frac{N}{(N-1)p(1-p)^{N-1}} \cdot T_\text{slot}.
\end{equation}
\end{prop}

\begin{NewProof}
Without loss of generality, we consider two adjacent refresh moments of a user A.
When there is a successful transmission, we denote the probabilities that the packet is from A and other users by $p_{A|S}$ and $p_{{\overline{A}}|S}$, respectively. An immediate result is that
$$p_{A|S} = \frac{1}{N},~~~p_{{\overline{A}}|S} = \frac{N-1}{N}.$$
Suppose the refresh time of user A contains $n_A$ consecutive packets successfully transmitted by A and $n_{\overline{A}}$ consecutive packets successfully transmitted by other users, where $n_A, n_{\overline{A}} \in \{1,2,3,\cdots\}$.
The probability that this transmission pattern occurs is $$p_{A|S}^{n_A} \cdot p_{{\overline{A}}|S}^{n_{\overline{A}}}=\left(\!\frac{1}{N}\!\right)^{n_A} \!\! \left(\!\frac{N-1}{N}\!\right)^{n_{\overline{A}}}.$$

The refresh time of user A can then be written as
\begin{equation*}
\mathcal{T}_A=\sum_{i=1}^{n_A} {T_{A,i}} +\sum_{j=1}^{n_{\overline{A}}} {T_{{\overline{A}},j}},
\end{equation*}
where $T_{A,i}$ ($T_{{\overline{A}},j}$) is the time it takes for user A (other users) to transmit the $i$-th ($j$-th) successful packet. 
We emphasize that $T_{A,i}$ and $T_{{\overline{A}},j}$ may not equal $T_\text{slot}$ because of collisions. 

Then, $\mathbb{E}[\mathcal{T}_A]$ is given by
\begin{eqnarray}\label{eq:III-7}
\mathbb{E}[\mathcal{T}_A] =&&\hspace{-0.6cm} 
\mathbb{E}_{n_A,n_{\overline{A}}} \left[ \mathbb{E} \left( \left. \sum_{i=1}^{n_A} {T_{A,i}} +\sum_{j=1}^{n_{\overline{A}}} {T_{{\overline{A}},j}} \right| n_A, n_{\overline{A}} \right) \right] \nonumber \\
=&&\hspace{-0.6cm} \sum_{n_A=1}^{\infty} \sum_{n_{\overline{A}}=1}^{\infty} (n_A \overline{T}_A + n_{\overline{A}} \overline{T}_{\overline{A}}) \cdot \left(\!\frac{1}{N}\!\right)^{n_A} \!\!\!\left(\!\frac{N-1}{N}\!\right)^{n_{\overline{A}}} \nonumber \\
=&&\hspace{-0.6cm} \frac{N^2}{N-1}\cdot \frac{\overline{T}_A+(N-1)\overline{T}_{\overline{A}}}{N},
\end{eqnarray}
where $\overline{T}_A$ and $\overline{T}_{\overline{A}}$ denote the average values of $\{T_{A,i}\}$ and $\{T_{{\overline{A}},j}\}$, respectively (they also represent the average time it takes to successfully transmit a packet of A and other users, respectively). As a result, $\frac{\overline{T}_A+(N-1)\overline{T}_{\overline{A}}}{N}$ is exactly the average time of a successful transmission, That is, $\frac{\overline{T}_A+(N-1)\overline{T}_{\overline{A}}}{N}=\overline{T}$.

It can be shown that $\mathbb{E}[\mathcal{T}_n]$ is the same for all users with slotted Aloha.
Substituting \eqref{eq:III-4} into \eqref{eq:III-7} yields \eqref{eq:III-5}.
\end{NewProof}

Next, we derive the average number of refresh time that one cycle time contains in slotted Aloha.

\begin{lem}\label{lem:sequence_permutation}
Consider $k$ consecutive successfully transmitted packets, the number of permutations that contain packets from $N$ different users is given by
\begin{equation}\label{eq:III-8}
\mathcal{P}_N(k) = \sum_{i=0}^N \binom{N}{i}(-1)^i(N-i)^k.
\end{equation}
\end{lem}

\begin{NewProof}
When $N=1$, i.e., there is only one user in the network, it is obvious that $\mathcal{P}_1(k)=1$ and \eqref{eq:III-8} satisfies.
For a network with $N+1$ users, we have the following recursive formula
\begin{equation}\label{eq:III-9}
\mathcal{P}_{N+1}(k) = \sum_{k^\prime=1}^{k-1} \binom{k}{k^\prime} \cdot \mathcal{P}_N(k).
\end{equation}
Eq. \eqref{eq:III-9} indicates that $\mathcal{P}_{N+1}(k)$ can be derived by choosing $k^\prime$ (out of $k$, $k^\prime=1,2,\cdots,k-1$) packets that are from $N$ different users and the remaining $k-k^\prime$ packets are from the $(N+1)$-th user.
Given \eqref{eq:III-9}, \eqref{eq:III-8} can be proven by mathematical induction.
\end{NewProof}

\begin{lem}\label{lem:sequence_contain}
Consider a time period that contains $\mathcal{I}$ consecutive refresh time of a user. We define $\mathcal{E_I}$ an event that all the $N$ users have transmitted at least one packet in this time period. Then,
\begin{equation}\label{eq:III-10}
\Pr(\mathcal{E_I}) = \frac{1}{(N\!-\!1)^{\mathcal{I}}}\sum_{i=0}^{N-1}\! \binom{N\!-\!1}{i}(-1)^i\left(\!\frac{N\!-\!1\!-\!i}{i+1}\!\right)^{\mathcal{I}}.
\end{equation}
\end{lem}

\begin{NewProof}
Consider any user A and its $\mathcal{I}$ consecutive refresh time.
We denote the numbers of packets transmitted from users other than A in these $\mathcal{I}$ refresh time by $k_1, k_2, \cdots, k_{\mathcal{I}}$.
Note that they are i.i.d. random variables in slotted Aloha.
To derive the distribution of $k_i, i=1,2,\cdots,\mathcal{I}$, we use the concept of number of inter-transmissions derived in \cite{1378897}, which is denoted by $N_I$ (see the definition in the introduction).

In slotted Aloha, the distribution of $N_I$ can be written as $P_{N_I,k}\triangleq \Pr (N_I=k) = \frac{1}{N}\left(1-\frac{1}{N}\right)^k$.
Then, the distributions of $k_i$ are given by
$\Pr(k_i=k)=\Pr(N_I=k|N_I>0) = \frac{1}{N-1}\left(\frac{N-1}{N}\right)^k$, and the joint distribution of $\{k_1, k_2, \cdots, k_{\mathcal{I}}\}$ can be written as
\begin{equation}\label{eq:III-11}
\Pr(k_1, k_2, \cdots, k_{\mathcal{I}}) = \frac{1}{(N\!-\!1)^{\mathcal{I}}} \left(\frac{N-1}{N}\right)^{\sum k},
\end{equation}
where $\sum k \triangleq k_1+ k_2+\cdots+k_{\mathcal{I}}$.

Given $k_1, k_2, \cdots, k_{\mathcal{I}}$, if all the $N$ users
have transmitted at least one packet in this time period, the $\sum k$ packets that are not from user A must from the other $N-1$ users, the probability of which is given by
$$\frac{\mathcal{P}_{N-1}(\sum k)}{(N-1)^{\sum k}},$$
where $\mathcal{P}_{N-1}(\sum k)$ is defined in \eqref{eq:III-8} and $(N-1)^{\sum k}$ represents the total number of permutations. As a result, $\Pr(\mathcal{E_I})$ can be derived as
\begin{eqnarray*}\label{eq:III-12}
\Pr(\mathcal{E_I}) =&&\hspace{-0.6cm} \sum_{k_1, k_2, \cdots, k_{\mathcal{I}}=1}^{\infty} \frac{\mathcal{P}_{N-1}(\sum k)}{(N-1)^{\sum k}} \cdot \Pr(k_1, k_2, \cdots, k_{\mathcal{I}}) \nonumber \\
=&&\hspace{-0.6cm} \sum_{k_1, k_2, \cdots, k_{\mathcal{I}}=1}^{\infty}
\frac{\sum_{i=0}^{N-1} \binom{N-1}{i}(-1)^i(N-1-i)^{\sum k}}{(N-1)^{\sum k}} \nonumber \\
&&\hspace{2.7cm} \times \frac{1}{(N\!-\!1)^{\mathcal{I}}} \left(\frac{N-1}{N}\right)^{\sum k} \nonumber \\
=&&\hspace{-0.6cm} \frac{1}{(N\!-\!1)^{\mathcal{I}}} 
\sum_{i=0}^{N-1} \binom{N-1}{i} (-1)^i \times \nonumber \\
&&\hspace{2cm} \sum_{k_1, k_2, \cdots, k_{\mathcal{I}}=1}^{\infty} \left(\frac{N-1-i}{N}\right)^{\sum k} \nonumber \\
=&& \hspace{-0.6cm} \frac{1}{(N\!-\!1)^{\mathcal{I}}}\sum_{i=0}^{N-1}\! \binom{N\!-\!1}{i}(-1)^i\left(\!\frac{N\!-\!1\!-\!i}{i+1}\!\right)^{\mathcal{I}}.
\end{eqnarray*}
Finally, we arrive at \eqref{eq:III-10}.
\end{NewProof}

Based on Lemmas~\ref{lem:sequence_permutation} and \ref{lem:sequence_contain}, we are ready to derive $\mathbb{E}[I_r]$.

\begin{prop}\label{thm:RRT_num_ALOHA}
The average number of refresh time that one cycle time contains in slotted Aloha is given by
\begin{equation}\label{eq:III-13}
    \mathbb{E}[I_r]=\frac{N-1}{N}(1+H_{N-1}),
\end{equation}
where $H_{N-1} \triangleq \sum_{i=1}^{N-1}\frac{1}{i}$ is the $(N-1)$-th harmonic number.
\end{prop}

\begin{NewProof}
We first derive the distribution of $I_r$.
For one cycle time of a user that contains $\mathcal{I}$ refresh time, we have from Definition~\ref{defi:cycle_time} that
\begin{itemize}    [leftmargin=0.45cm]
    \item  All the $N$ users have transmitted at least one packet in the time period that contains $\mathcal{I}$ refresh time.
    \item If $\mathcal{I}>1$, the first $(\mathcal{I}-1)$ refresh time does not contain packets from all the $N$ users.
\end{itemize}

Therefore, we have
\begin{equation}\label{eq:III-14}
\Pr(I_r=\mathcal{I}) = 
\begin{cases}
    \Pr(\mathcal{E}_1),&\text{if}~\mathcal{I}=1;\\
    \Pr(\overline{\mathcal{E}_{\mathcal{I}-1}},\mathcal{E_I}) \\ 
    \hspace{0.6cm}=\Pr(\mathcal{E_I} - \Pr(\mathcal{E}_{\mathcal{I}-1},\mathcal{E_I}) \\
    \hspace{0.6cm}\overset{(a)}{=}\Pr(\mathcal{E_I}) - \Pr(\mathcal{E}_{\mathcal{I}-1}), & \text{if}~\mathcal{I}>1;
\end{cases}
\end{equation}
where $\mathcal{E_I}$ is the event defined in Lemma~\ref{lem:sequence_contain}, $\overline{\mathcal{E}_{\mathcal{I}-1}}$ is the complement of the event $\mathcal{E}_{\mathcal{I}-1}$, and $(a)$ follows because if the first $\mathcal{I}-1$ refresh time contains packets from all the $N$ users, then all the $\mathcal{I}$ refresh time also contains packets from $N$ users.

The distribution of $I_r$ in \eqref{eq:III-14} gives us
\begin{eqnarray}\label{eq:III-15}
\mathbb{E}[I_r]=&&\hspace{-0.6cm} \sum_{I=1}^{\infty} I\cdot\Pr(\mathcal{E}_I) \\
=&&\hspace{-0.6cm} 1+\frac{1}{N}\sum_{i=1}^{N-1} \binom{N\!-\!1}{i} (-1)^i \left( 1-\frac{N-1}{i} \right) \nonumber .
\end{eqnarray}
We further have
\begin{eqnarray*}
&&\hspace{-0.6cm}\sum_{i=1}^{N-1} \binom{N\!-\!1}{i} (-1)^i= -1,\\
&&\hspace{-0.6cm}\sum_{i=1}^{N-1} \binom{N\!-\!1}{i} \frac{(-1)^i}{i} = -\sum_{i=1}^{N-1} \frac{1}{i} = -H_{N-1},
\end{eqnarray*}
where the latter can be proven by mathematical induction. Overall, \eqref{eq:III-15} can be refined as
\begin{equation*}
\mathbb{E}[I_r]=1+\frac{1}{N} [ -1+(N-1)H_{N-1} ]
= \frac{N-1}{N}(1+H_{N-1}),
\end{equation*}
completing the proof.
\end{NewProof}

\begin{thm}\label{thm:CCT_ALOHA}
The CCT of slotted Aloha is
\begin{equation}\label{eq:III-16}
    \Psi_\text{slotted-Aloha}=\frac{1+\sum_{i=1}^{N-1}\frac{1}{i}}{p(1-p)^{N-1}}\cdot T_\text{slot}.    
\end{equation}
\end{thm}

\begin{NewProof}
The average cycle time of the $n$-th user can be obtained by substituting \eqref{eq:III-5} and \eqref{eq:III-13} into \eqref{eq:III-3}, yielding 
\begin{eqnarray*}
    \mathbb{E}[\Gamma_n]=&&\hspace{-0.6cm} \frac{N-1}{N}(1+H_{N-1}) \times \frac{N}{(N-1)p(1-p)^{N-1}} T_\text{slot} \\
    =&&\hspace{-0.6cm} \frac{1+H_{N-1}}{p(1-p)^{N-1}} T_\text{slot}.
\end{eqnarray*}
Note that $\mathbb{E}[\Gamma_n]$ is same for all users. As per Definition~\ref{defi:channel_cycle_time}, the CCT of slotted Aloha is exactly $\mathbb{E}[\Gamma_n]$.
\end{NewProof}

\subsection{CCT-optimal slotted Aloha}\label{sec:III-B}
Given the closed-form channel cycle time, we next investigate the CCT-optimal slotted Aloha. That is, we optimize the transmission probability $p$ to obtain the short-term fairest slotted Aloha.

Differentiating \eqref{eq:III-16} with respect to $p$ and setting the results to 0, it is easy to find that the optimal $p^*=\frac{1}{N}$, in which case slotted Aloha has the minimum channel cycle time:
\begin{equation}\label{eq:III-17}
\Psi^*_\text{slotted-Aloha} = \frac{N(1+H_{N-1})}{(1-\frac{1}{N})^{N-1}} T_\text{slot} = \frac{N(1+H_{N-1})}{(1-\frac{1}{N})^{N-1}} \ell_\text{pkt}.
\end{equation}

It is worth noting that the throughput of slotted Aloha reaches the maximum when the average number of transmission trials per slot $G=1$.
For the network with $N$ users, $G$ can be calculated as
$$G=\sum_{i=1}^N i\cdot \binom{N}{i} p^i (1-p)^{N-i}=Np.$$
Therefore, for slotted Aloha, the transmission probability that achieves the minimum CCT also gives us the maximum throughput.

In contrast, other short-term metrics fail to offer such valuable insights. The fundamental reason for this lies in the fact that CCT also captures the transmission delay and jitter, establishing an intrinsic connection with throughput. Other metrics solely focus on successfully transmitted packets without any regard for time, constraining their capacity to comprehensively characterize the short-term behavior of a MAC protocol.

\section{CCT of CSMA/CA}\label{sec:IV}
CSMA/CA is the MAC protocol widely used in Wi-Fi networks \cite{9502043}.
In CSMA/CA, each terminal carrier senses the channel before transmission and retransmits after a binary exponential backoff in the case of transmission failures. That is, the average backoff time is doubled after a transmission failure to reduce the collision probability.

\begin{figure}[t]
\centering
\includegraphics[width=0.9\columnwidth]{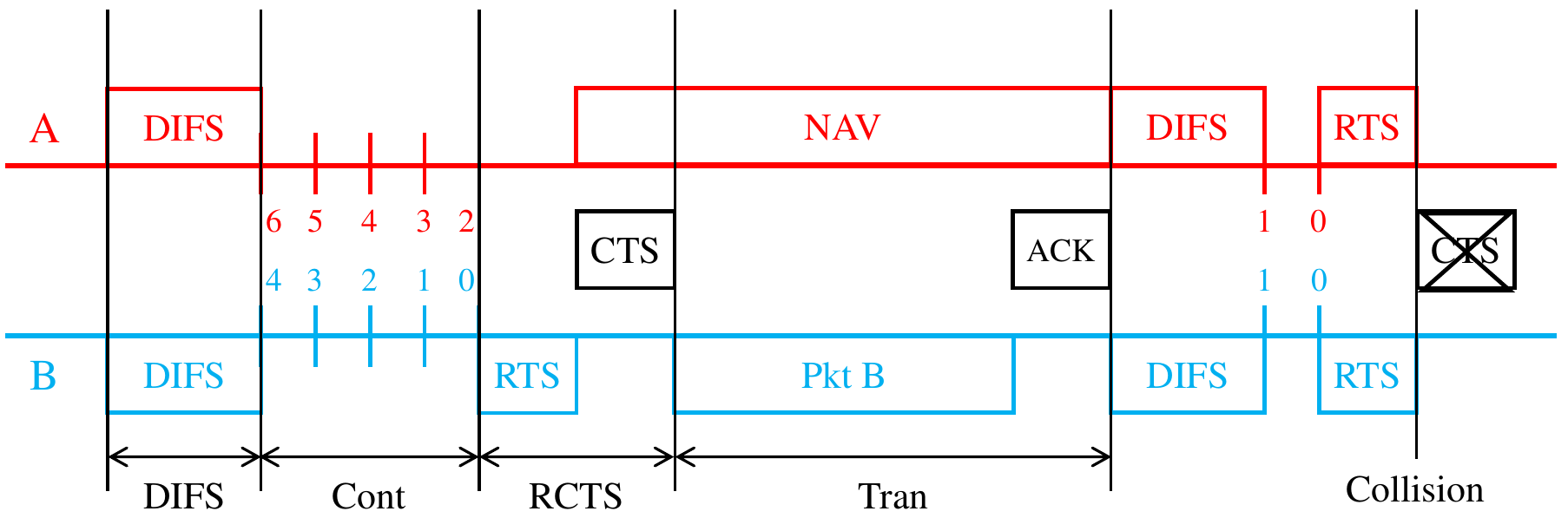}
\caption{An example of the CSMA/CA protocol with a successful transmission of user B and a collision.}
\label{fig:CSMA_rcts_actions}
\end{figure}

The analysis of CCT for CSMA/CA is more challenging.
To facilitate exposition, we first briefly review the CSMA/CA protocol using the concrete example given in Fig.~\ref{fig:CSMA_rcts_actions} with two users A and B in the network.
As shown, there is a successful transmission of user B followed by a collision.
\begin{itemize} [leftmargin=0.45cm]
    \item In the beginning, the channel is idle. After a distributed inter-frame space (DIFS), both users enter the contention process and select a random counter according to its current contention window.
    \item As shown in Fig.~\ref{fig:CSMA_rcts_actions}, the counter of user B is smaller, hence it occupies the channel by sending a Request to Send (RTS) packet when the countdown goes to zero. Upon receiving the RTS packet, user A updates its network allocation vector (NAV) and suspend the contention process (the counter will be frozen). User B, on the other hand, starts to transmit the data packet after receiving the AP's Clear to Send (CTS) packet.
    \item When the transmission of user B finishes, both users enter another DIFS and contention process. This time, the randomly generated counter of user B equals the remaining counter of user A. As a consequence, the countdown of both users ends at the same time and a collision happens. Since no CTS is received, both users detect the collision. They double their contention windows and select new backoff counters based on the updated contention windows.
\end{itemize}

Compared to slotted Aloha, the analysis and evaluation of CSMA/CA pose more complex challenges. This stems from the fact that successful transmissions in CSMA/CA are considerably interrelated, in contrast to the independent transmissions in slotted Aloha. Consequently, a distinct approach must be adopted to dissect the CCT of CSMA/CA.

To gain analytical insights and derive closed-form expressions, this section will concentrate on a network featuring two users, designated as A and B. 
For such a two-user network, the concepts of cycle time and refresh time become interchangeable, yielding
\begin{equation}\label{eq:IV-0}
\mathbb{E}[\Gamma_n]=\mathbb{E}[ \mathcal{T}_n].
\end{equation}
With this property and Corollary \ref{thm:cor1}, we will meticulously analyze CSMA/CA in both the RTS/CTS mode and the basic mode (without RTS/CTS).
For networks involving more users, we will resort to simulations to evaluate their channel cycle time.

\subsection{CSMA/CA in the RTS/CTS mode}\label{sec:IV-A}

\begin{figure}[t]
\centering
\includegraphics[width=0.9\columnwidth]{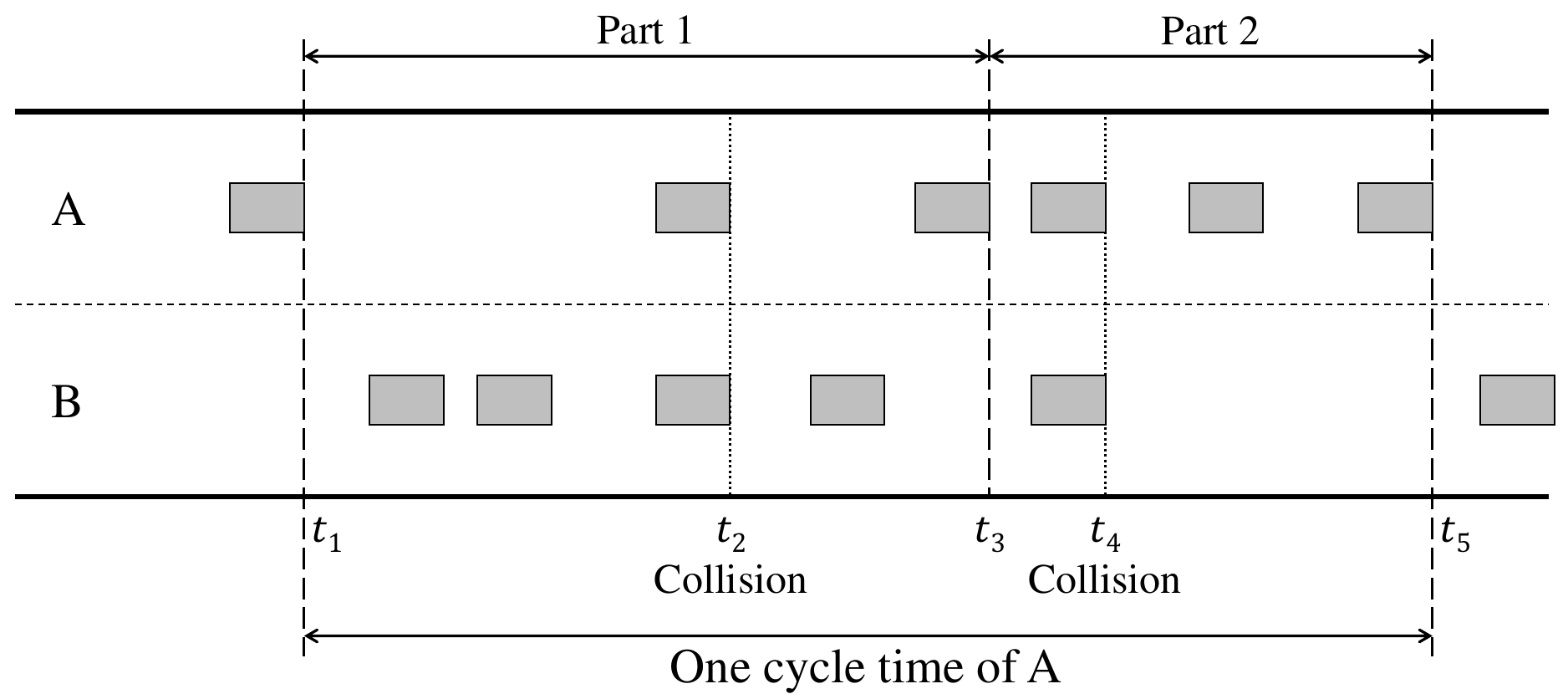}
\caption{In CSMA/CA, the cycle time of a user can be divided into two parts.}
\label{fig:CSMA_two_parts}
\end{figure}


When operated with CSMA/CA, an important observation is that the cycle time of any user can be divided into two parts.
Let us consider the example in Fig.~\ref{fig:CSMA_two_parts}, where user A's cycle time is divided into Part 1 and Part 2. Specifically,
\begin{enumerate} [leftmargin=0.5cm]
    \item Part 1 consists of all transmissions of user B and the first successful transmission of user A. 
    We denote by $n_B$ the number of successful transmissions of user B in Part 1.
    \item Part 2 consists of all remaining successful transmissions of user A, and we denote its number by $n_A^\prime$. Note that $n_A^\prime$ can be 0, in which case the duration of Part 2 is 0.
\end{enumerate}

In Fig.~\ref{fig:CSMA_two_parts}, the duration of Part 1 and Part 2 is $t_3-t_1$ and $t_5-t_3$, respectively. 

\begin{lem}\label{lem:nAB}
In CSMA/CA, users have the same distributions of the number of inter-transmissions. Therefore, we can denote the number of inter-transmissions by a common random variable $N_I$.
Let $P_{N_I,k}\triangleq \Pr (N_I=k)$. The distribution of $n_B$ and $n_A^\prime$ are given by 
\begin{eqnarray}\label{eq:IV-1}
&&\hspace{-0.6cm} \Pr (n_B=k) = \Pr (N_I=k|N_I>0) = \frac{P_{N_I,k}}{1-P_{N_I,0}}, \\
\label{eq:IV-1b}
&&\hspace{-0.6cm} \Pr (n_A^\prime=k) = P_{N_I,0}^k (1-P_{N_I,0}).
\end{eqnarray}
In particular,
\begin{equation}\label{eq:IV-2}
\mathbb{E}(n_B) = \frac{\mathbb{E}(N_I)}{1-P_{N_I,0}},~~
\mathbb{E}(n_A^\prime) = \frac{P_{N_I,0}}{1-P_{N_I,0}}.
\end{equation}
\end{lem}

\begin{NewProof}
    According to the definition, $n_B$ is the number of inter-transmissions $N_I$ under the condition that $N_I>0$. Eq.~\eqref{eq:IV-1} follows. On the other hand, for $n_A^\prime$, it can be viewed that there are $n_A^\prime$ transmissions with $N_I=0$ followed by a transmission with $N_I>0$, thus we have \eqref{eq:IV-1b}.
\end{NewProof}

Next, we derive the duration of Parts 1 and 2, respectively. To ease exposition, we introduce the following notations:
\begin{itemize}[leftmargin=0.45cm]
\item We use $\ell$ to represent the duration of a packet or a process, e.g., $\ell_\text{difs}$, $\ell_\text{pkt}$, $\ell_\text{ack}$, $\ell_\text{rts}$, $\ell_\text{cts}$.
We further define $\ell_\text{tran}\triangleq\ell_\text{pkt}+\ell_\text{ack}$,
$\ell_\text{rcts}\triangleq\ell_\text{rts}+\ell_\text{cts}$, and
$\ell_\text{nav}\triangleq\ell_\text{tran}+\ell_\text{rcts}-T_\text{slot}$, where $T_\text{slot}$ is the slot time of CSMA/CA.
\item For each user, the minimum contention window size is $\text{CW}_\text{min}$. After the $i$-th collision, the contention window size $\text{CW}_i = \min\{2^i\cdot \text{CW}_\text{min}, \text{CW}_\text{max}\}$, where $\text{CW}_\text{max}=2^{\beta}\cdot \text{CW}_\text{min}$.
The random backoff counter $\lambda_i$ is uniformly chosen from $[1,\text{CW}_i]$, i.e., $\lambda_i\sim U(1,\text{CW}_i)$.
\end{itemize}

\begin{lem}\label{lem:part1}
The duration of Part 1 is
\begin{equation}\label{eq:IV-3}
T_\text{part1} = n_B (\ell_\text{difs} + \ell_\text{nav}) + \ell_\text{tran} +  \sum_{i=0}^{\rho_1} (\ell_\text{difs} + \ell_\text{rcts} + \lambda_i),
\end{equation}
where $\rho_1$ is the number of collisions in Part 1.
\end{lem}

\begin{NewProof}
See Appendix \ref{sec:AppA}.
\end{NewProof}

\begin{lem}\label{lem:part2}
The duration of Part 2 is
\begin{equation}\label{eq:IV-5}
    T_\text{part2} = \sum_{j=0}^{n_A^\prime} \left[ \ell_\text{tran} + \sum_{i=0}^{\rho_2^{(j)}} \left( \ell_\text{difs} + \ell_\text{rcts} + \lambda_i^{(j)} \right) \right],
\end{equation}
where $\rho_2^{(j)}$ is the number of collisions that occurred in Part 2 for the $j$-th transmission of user A, and $\lambda_i^{(j)}$ is the random backoff counter after the $i$-th collision for the $j$-th transmission of A. We emphasize that $\rho_2^{(j)}$ are independent and identically distributed (i.i.d.) random variables, $\forall j$. 
\end{lem}

\begin{NewProof}
See Appendix \ref{sec:AppA}.
\end{NewProof}

Based on Lemmas~\ref{lem:nAB}, \ref{lem:part1}, and \ref{lem:part2}, we are ready to derive the CCT of CSMA/CA in the RTS/CTS mode.

\begin{thm}\label{prop:CCT_CSMA_basic}
Consider a two-user multiple-access network. The CCT of CSMA/CA in the RTS/CTS mode is given by
\begin{eqnarray}\label{eq:IV-7}
    \Psi_\text{CSMA/CA} =&&\hspace{-0.6cm} \frac{1}{1-P_{N_I,0}} \!\left[  \ell_\text{difs} + \ell_\text{nav} + \ell_\text{tran}
    \!+\!  \frac{\ell_\text{difs} \!+\! \ell_\text{rcts}}{1-p_c}  \right.  \\
    &&\hspace{-0.5cm} + \! \left. \sum_{i=0}^{\beta-1} p_c^i \frac{1+2^i\text{CW}_\text{min}}{2} + \frac{p_c^{\beta}}{1-p_c} \frac{1+2^{\beta}\text{CW}_\text{min}}{2}    
    \right], \nonumber
\end{eqnarray}
where $p_c$ is the collision probability of CSMA/CA that satisfies the following recursive equation with a unique solution:
\begin{equation}\label{eq:IV-10}
p_c = \frac{2(1-2p_c)}{(1-2p_c)(\text{CW}_\text{min}+3)+p_c\text{CW}_\text{min}[1-(2p_c)^{\beta}]}.
\end{equation}
\end{thm}

\begin{NewProof}
See Appendix \ref{sec:AppB}.
\end{NewProof}

In the context of networks involving more than two users, the derivation of cycle time for each user becomes significantly more intricate.
As outlined in \eqref{eq:III-2}, the examination of a network with an arbitrary number of users hinges upon our understanding of the number of refresh time encapsulated within one cycle time, i.e., the random variable $I_r$.
In the simple scenario of two users, we have $I_r=1$, lending itself to a more manageable analysis. However, in cases where $I_r$ takes on a more general value greater than 1, the successful transmissions occurring between consecutive refresh time exhibit substantial interdependence (that is, $I_r$ becomes non-independent from $\mathcal{T}_n$), posing challenges in deducing the distribution of $I_r$.
Later in Section \ref{sec:VI}, we will perform a simulation-based exploration of CCT for CSMA/CA in scenarios involving more users.

\subsection{CCT-optimal CSMA/CA in the RTS/CTS mode}\label{sec:IV-D}
Assuming $\text{CW}_\text{max}=\text{CW}_\text{min}$, our focus is on identifying the optimal CW value to optimize the CCT of CSMA/CA.

From \eqref{eq:IV-7}, we have
\begin{equation*}\label{eq:IV-11}
    \Psi_\text{CSMA/CA} \!=\!
    \frac{1}{1 \!-\! P_{N_I,0}} \!\! \left[ \frac{2(\ell_\text{difs} \!+\! \ell_\text{rcts})}{\text{CW}_\text{min}+1} \!+\! \frac{\text{CW}_\text{min}+1}{2} \!+\!C \right],
\end{equation*}
where $C \triangleq 2\ell_\text{difs} + \ell_\text{nav} + \ell_\text{rcts} + \ell_\text{tran} +1$.
It is easy to find that the minimum $\Psi_\text{CSMA/CA}^*$ is obtained when $\text{CW}_\text{min}=2\sqrt{\ell_\text{difs} + \ell_\text{rcts}}-1$.

On the other hand, \cite{840210} shows that, when $\text{CW}_\text{max}=\text{CW}_\text{min}$, the maximum throughput of CSMA/CA in a two-user multiple access network can be obtained when $\text{CW}_\text{min} = 2\sqrt{\ell_\text{difs} + \ell_\text{tran}}-1$. Therefore, we have the same result as that in slotted Aloha: the parameters that achieve the minimum channel cycle time also give us the maximum throughput.

\subsection{CSMA/CA in the basic mode}\label{sec:IV-C}

The RTS/CTS mechanism prevents the hidden terminal problem in CSMA/CA by enabling devices to signal their intent to transmit and receive data.
While it does offer benefits, this mechanism also brings in extra handshake overhead, especially when dealing with short packet lengths.
As a result, many Wi-Fi routers available today come with the option to deactivate the RTS/CTS mechanism and shift into the basic CSMA/CA mode. In such scenarios, when the countdown reaches zero, a user can directly transmit a data packet, bypassing the RTS/CTS handshake.

Consider a two-user multiple-access network operated with the basic mode of CSMA/CA.
If the two users are not hidden terminals to each other, the CCT can be derived in a similar way as Theorem~\ref{prop:CCT_CSMA_basic}, giving
\begin{eqnarray}\label{eq:IV-12}
    \Psi_\text{CSMA/CA-Basic} \!=\! &&\hspace{-0.6cm} \frac{1}{1\!-\!P_{N_I,0}} \!\! \left[ \ell_\text{difs} \!+\! \ell_\text{tran} - T_\text{slot} 
    \!+\!  \frac{\ell_\text{difs} \!+\! \ell_\text{tran}}{1-p_c} \right. \nonumber \\ 
    &&\hspace{-2.2cm} + \left. \sum_{i=0}^{\beta-1} p_c^i \frac{1+2^i\text{CW}_\text{min}}{2} + \frac{p_c^{\beta}}{1-p_c} \frac{1+2^{\beta}\text{CW}_\text{min}}{2} \right].
\end{eqnarray}

However, the scenario takes a different turn when the two users are hidden terminals to each other. In this case, the channel appears consistently idle to both users. Consequently, the contention process of one user remains unaffected even when the other user is transmitting packets.
The suspension of the contention process occurs only upon receiving an ACK frame from the access point indicating the successful transmission of the other user.

To derive the CCT of CSMA/CA in this case, we adopt the same approach as Section \ref{sec:IV-A} and divide the cycle time of each user into two parts.
Following Lemmas \ref{lem:part1} and \ref{lem:part2}, the duration of Parts 1 and 2 can be derived as follows:
\begin{eqnarray*}\label{eq:V-1}
&&\hspace{-0.6cm} T_\text{part1-hidden} = n_B \ell_\text{difs} + \sum_{i=0}^{\rho_1} (\ell_\text{difs} + \ell_\text{tran} + \lambda_i),\\
&&\hspace{-0.6cm} T_\text{part2-hidden} = \sum_{j=0}^{n_A^\prime} \sum_{i=0}^{\rho_2^{(j)}} \left( \ell_\text{difs} + \ell_\text{tran} + \lambda_i^{(j)} \right).
\end{eqnarray*}

\begin{thm}\label{prop:CCT_CSMA_hidden}
Consider a two-user multiple-access network, where the two users are hidden terminals to each other. Under the basic mode of CSMA/CA, the CCT of the system is given by
\begin{eqnarray}\label{eq:V-2}
    \Psi_\text{CSMA/CA-hidden} = &&\hspace{-0.6cm} \\
    &&\hspace{-2cm} \frac{1}{1-P_{N_I,0}}\left[ \ell_\text{difs} \cdot \mathbb{E}(N_I) + 
    \mathbb{E}(n_c) \cdot (\ell_\text{difs} + \ell_\text{tran}) + \mu 
    \right], \nonumber
\end{eqnarray}
where $\mathbb{E}(n_c)$ is the average number of collisions it takes to successfully transmit a packet,
\begin{eqnarray*}\label{eq:V-3}
    &&\hspace{-0.6cm} \mathbb{E}(n_c) = 
    1+\sum_{k=1}^{m-1}\prod_{i=0}^{k-1}p_{c,i} + \frac{1}{1-p_{c,m}}\prod_{i=0}^{m-1}p_{c,i},\\
    &&\hspace{-0.6cm} \mu =
    \mathbb{E}(\lambda_0) + \sum_{k=1}^{m-1}\mathbb{E}(\lambda_k) \! \prod_{i=0}^{k-1}p_{c,i} + \frac{\mathbb{E}(\lambda_m)}{1-p_{c,m}} \! \prod_{i=0}^{m-1}p_{c,i},
\end{eqnarray*}
and $p_{c,i}$, $i=0,1,\cdots,m$, is the collision probability in the $i$-th backoff stage, and $\mathbb{E}(\lambda_i)=\frac{1+2^i\text{CW}_\text{min}}{2}$.

\end{thm}

\begin{NewProof}
The proof follows Theorem \ref{prop:CCT_CSMA_basic} in general and is omitted.
\end{NewProof}

Computing $\Psi_\text{CSMA/CA-hidden}$ from \eqref{eq:V-2} is not easy, mainly due to the difficulty in deriving the collision probabilities $p_{c,i}$ for each backoff stage. To provide a more illustrative perspective on this issue, we can describe the state transition of a specific user in the CSMA/CA system as a Markov chain. As depicted in Figure 4, this Markov chain's state $(i,j)$ consists of two components: the backoff stages $i$ and the current counter values $j$. In the counter down phase, the Markov chain transits to the left; in the case of a collision, the Markov chain transits to the next layer; in the case of a successful transmission, the Markov chain transits back to the first layer.

In \eqref{eq:IV-7} and \eqref{eq:IV-12}, there are no hidden terminals and the collision probabilities at different backoff stages are roughly equal, that is $ p_{c,0}\approx p_{c,1}\approx \cdots \approx p_{c,m}\approx p_{c}$. In this case, $p_{c}$ can be calculated using the recursive expression provided in \eqref{eq:IV-10}. However, the situation changes when two users are hidden terminals to each other, causing collision probabilities to become dependent on the backoff stage. This complexity renders the computation of $\mathbb{E}(n_c)$ and $\mu$ a formidable challenge.

To tackle this challenge, we next present an alternative representation of $\mathbb{E}(n_c)$ and $\mu$ using the concept of global collision probability. This will consequently provide an upper bound for CCT.

\begin{figure}[t]
\centering
\includegraphics[width=0.95\columnwidth]{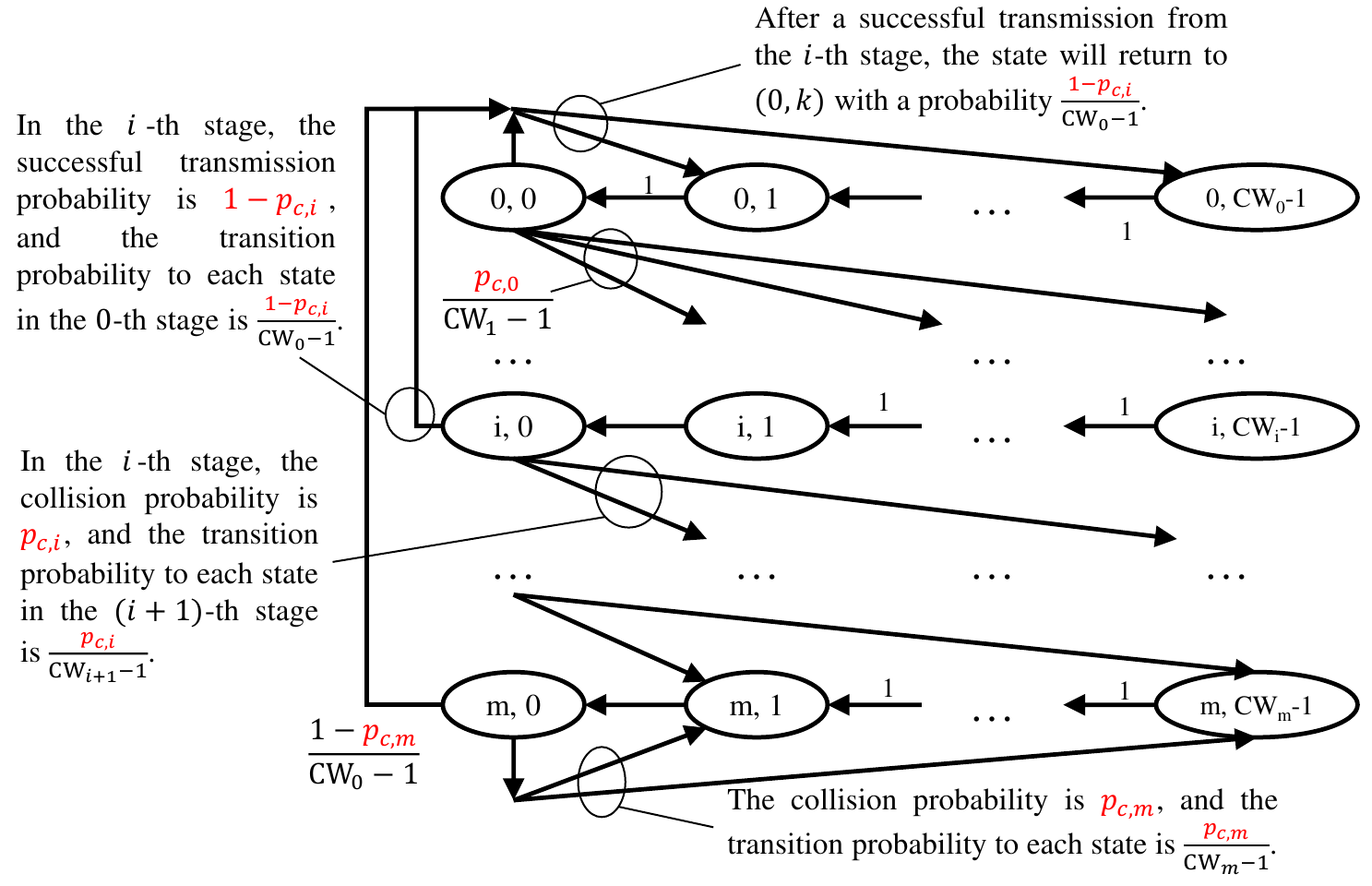}
\caption{The Markov Chain model for the user state transitions in a two-user network operated with the basic mode of CSMA/CA. The two users are hidden terminals; hence, the collision probabilities are dependent on the backoff stages.}
\label{fig:CSMA_Markov_hidden}
\end{figure}

\begin{lem}\label{thm:coll_prob_CSMA_hidden}
Let $q_c$ be the global collision probability of CSMA/CA, i.e., the collision probability of a packet irrespective of the current backoff stage\footnote{When there is no hidden terminal, we have $q_c=p_c$, as derived in Theorem \ref{prop:CCT_CSMA_basic}. On the other hand, in the presence of hidden terminals, $q_c$ can be obtained by the methods given in \cite{4542784,4799015}.}. The average number of collisions it takes to
successfully transmit a packet $\mathbb{E}(n_c)$ can be expressed as
\begin{equation}\label{eq:V-4}
    \mathbb{E}(n_c) = \frac{1}{1-q_c}.
\end{equation}
\end{lem}

\begin{NewProof}
Assuming that there are in total $a_i$ transmissions in the $i$-th backoff stage, and $b_i$ out of these $a_i$ transmissions are failed. We then have
\begin{eqnarray*}\label{eq:V-6}
    &&\hspace{-0.6cm} p_{c,i}=\frac{b_i}{a_i},i=0,1,\cdots,m,\\
    &&\hspace{-0.6cm} q_c=\frac{\sum_{i=0}^m b_i}{\sum_{i=0}^m a_i}.
\end{eqnarray*}

To show Lemma \ref{thm:coll_prob_CSMA_hidden} is true, we only need to prove
\begin{equation}\label{eq:V-5}
    1+\sum_{k=1}^{m-1}\prod_{i=0}^{k-1}p_{c,i} + \frac{1}{1-p_{c,m}}\prod_{i=0}^{m-1}p_{c,i}=\frac{1}{1-q_c}.
\end{equation}

Note that in stage $i$, $0\le i < m$, all the $b_i$ failed transmissions will transit to the $(i+1)$-th stage. That is to say, the number of failed transmissions in stage $i$ equals the total number of transmissions in stage $i+1$, except for $i=m-1$, in which case the failed transmissions must be successful in stage $m$. As a result, we have
\begin{eqnarray*}\label{eq:V-7}
    &&\hspace{-0.6cm} b_i=a_{i+1},i=0,1,\cdots,m-2,\\
    &&\hspace{-0.6cm} b_{m-1}=a_m-b_m.
\end{eqnarray*}

The left hand side of \eqref{eq:V-5} can be further refined as
\begin{eqnarray*}\label{eq:V-8}
    &&\hspace{-0.6cm} 1+\sum_{k=1}^{m-1}\prod_{i=0}^{k-1}p_{c,i} + \frac{1}{1-p_{c,m}}\prod_{i=0}^{m-1}p_{c,i}\\
    =&&\hspace{-0.6cm} 1+\sum_{k=1}^{m-1}\prod_{i=0}^{k-1}\frac{b_i}{a_i} + \frac{1}{1-\frac{b_m}{a_m}}\prod_{i=0}^{m-1}\frac{b_i}{a_i} \\
    =&&\hspace{-0.6cm} 1+\sum_{k=1}^{m-1}\frac{b_{k-1}}{a_0} + \frac{a_m}{a_m-b_m}\frac{b_{m-1}}{a_0} \\
    =&&\hspace{-0.6cm} 1+\frac{1}{a_0} \sum_{k=0}^{m}b_m=\frac{\sum_{i=0}^m a_i}{a_0}. \\
\end{eqnarray*}
On the other hand, we have
\begin{equation*}\label{eq:V-9}
    \frac{1}{1-q_c} = \frac{\sum_{i=0}^m a_i}{\sum_{i=0}^m a_i-\sum_{i=0}^m b_i} = \frac{\sum_{i=0}^m a_i}{a_0},
\end{equation*}
proving the lemma.
\end{NewProof}

\begin{thm}\label{thm:CW_bounds_CSMA_hidden}
Consider a two-user multiple-access network,
where the two users are hidden terminals to each other. Under the basic mode of CSMA/CA, the CCT of the system is upper bounded by
\begin{eqnarray}\label{eq:V-10}
    \Psi_\text{CSMA/CA-hidden}\leq&&\hspace{-0.6cm} \frac{1}{1\!-\!P_{N_I,0}} \!\times\! \left\{ \ell_\text{difs} \!\cdot\! \mathbb{E}(N_I) \!+\! \frac{\ell_\text{difs} \!+\! \ell_\text{tran}}{1-q_c} \right. \nonumber \\
    &&\hspace{-1.2cm}  \left. + \frac{1}{2(1\!-\!q_c)} \!+\! \frac{\text{CW}_\text{min}}{2}\!\left[1\!+\!\frac{2^m q_c}{1-q_c}\right] 
    \right\}.
\end{eqnarray}

\end{thm}

\begin{NewProof}
According to Theorem \ref{prop:CCT_CSMA_hidden},
\begin{eqnarray*}
    \mu &&\hspace{-0.6cm} =
    \frac{1+\text{CW}_\text{min}}{2} + \sum_{k=1}^{m-1} \frac{1+2^k\text{CW}_\text{min}}{2} \! \prod_{i=0}^{k-1}p_{c,i} \\
    &&\hspace{1.5cm} + \frac{1+2^m\text{CW}_\text{min}}{2(1-p_{c,m})} \! \prod_{i=0}^{m-1}p_{c,i},\\
    &&\hspace{-0.8cm}= \frac{1}{2} \mathbb{E}(n_c) \!+\!     \frac{\text{CW}_\text{min}}{2} \! \left[\!1\!+\!\sum_{k=1}^{m-1} 2^k \! \prod_{i=0}^{k-1}p_{c,i}\!+\!\frac{2^m}{1\!-\!p_{c,m}} \! \prod_{i=0}^{m-1}p_{c,i}\!\right] .
\end{eqnarray*}

Further, we have
\begin{eqnarray}\label{eq:v-12}
&&\hspace{-0.2cm} 1+\sum_{k=1}^{m-1} 2^k \prod_{i=0}^{k-1}p_{c,i}+\frac{2^m}{1-p_{c,m}} \prod_{i=0}^{m-1}p_{c,i} \nonumber\\
&&\hspace{-0.6cm} \le \frac{1}{1\!-\!q_c}\!+\!\!\left(\frac{1}{1 \!-\!q_c}\!-\!1\right) \!\! \left[1+(2^2-2)+\cdots+(2^m-2^{m-1})\right] \nonumber\\
&&\hspace{-0.6cm} = 1+2^m\frac{q_c}{1-q_c},
\end{eqnarray}
where the equation holds when $m\in\{0,1\}$.

Substituting \eqref{eq:V-4} and \eqref{eq:v-12} into \eqref{eq:V-2}, we arrive at \eqref{eq:V-10}.
\end{NewProof}


\section{Optimal CCT in Heterogeneous Networks}\label{sec:V}

Amidst the rapid advancements in deep learning technology, people have embarked on a journey to harness the potential of deep reinforcement learning (DRL) techniques for creating random access protocols for WiFi networks \cite{shao2022learning,8665952,8902705}. This exciting convergence of DRL and multiple access protocols, known as deep reinforcement learning-based multiple access (DRLMA), leads to many opportunities. Here, intelligent users take center stage, autonomously tuning and perfecting their transmission policies in various scenarios. The remarkable prowess of DRLMA lies in its unsupervised nature, versatility, adaptability, and predictive capacities. Leveraging DRL, each user dynamically adapts to the behaviors of others based on their past interactions, culminating in a repertoire of multi-access policies.

Within the realm of this research, a particularly intriguing scenario emerges: the heterogeneous network. In its basic form, the heterogeneous network envisions an intelligent user coexisting with a conventional WiFi user within the same basic service set (BSS). The ultimate challenge rests in empowering the intelligent node to acquire the optimal transmission policy to optimize a given criterion.
In this section, our focus pivots to an interesting problem: if we consider the CCT as the ultimate optimization goal, what is the minimum achievable CCT for the basic two-user heterogeneous network?

\subsection{DLRMA versus the RTS/CTS mode CSMA/CA}\label{sec:V-A}
We first consider the case where the CSMA/CA user is in the RTS/CTS mode.

\begin{prop}\label{lem:optimal_CSMA_DRL}
Consider a two-user heterogeneous network, where one user operates with the RTS/CTS mode CSMA/CA and the other user operates with DRLMA.
The minimum achievable CCT of the network is given by
\begin{equation}\label{eq:V-17}
    \Psi^*_\text{DRLMA} = \ell_\text{difs} + \ell_\text{rcts} + 2\ell_\text{tran} + \frac{1+\text{CW}_\text{min}}{2}.
\end{equation}
\end{prop}

\begin{figure}[t]
\centering
\includegraphics[width=0.9\columnwidth]{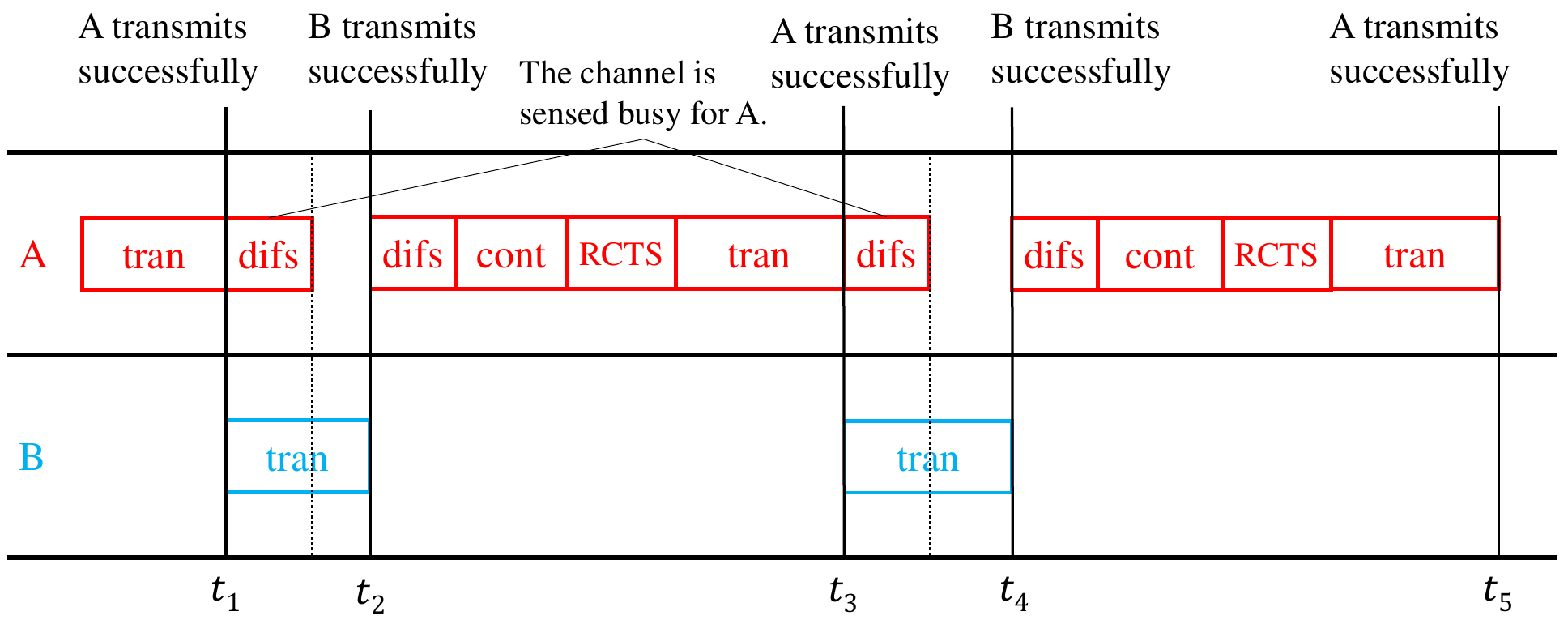}
\caption{The optimal transmission policy of the DRLMA user in Proposition \ref{lem:optimal_CSMA_DRL}.}
\label{fig:CSMA_DRL_optimal}
\end{figure}

\begin{NewProof}
(sketch) Let us denote the two users by A and B, respectively, where user A operates with the RTS/CTS mode CSMA/CA and user B operates with DRLMA.

With RTS/CTS, user A can sense any transmission within the BSS. Thus, the contention process of user A will be suspended once B starts to transmit. It can be shown that the optimal transmission policy of user B that leads to the minimum CCT is given by
\begin{enumerate}
\item User B stays silent until an ACK is broadcasted by the AP, indicating a successful transmission of user A.
\item User B immediately transmits a packet, then goes to 1).
\end{enumerate}

Intuitively, there should be one successful transmission of user B between the two successful transmissions of user A. The above transmission policy of user B ensures that the cycle time of both A and B is minimized.

Fig.~\ref{fig:CSMA_DRL_optimal} depicts the optimal transmission patterns of users A and B under the above policy. The cycle time of users A and B can be written as
$$\Gamma_A=\Gamma_B=\ell_\text{tran} + (\ell_\text{difs} + \ell_\text{rcts} + \ell_\text{tran} + \lambda_0),$$
where $\lambda_0\sim U(1,\text{CW}_0)$. The CCT can be derived accordingly.

\end{NewProof}




\subsection{DLRMA versus the basic mode CSMA/CA}\label{sec:V-B}
Next, we consider the case where the CSMA/CA user is in the basic mode. If the CSMA/CA user can sense the transmission of the DRLMA user (i.e., they are not hidden terminals), the optimal CCT can be derived in a similar way as Proposition \ref{lem:optimal_CSMA_DRL}, yielding
\begin{equation}\label{eq:V-18}
    \Psi^*_\text{DRLMA} = \ell_\text{difs} + 2\ell_\text{tran} + \frac{1+\text{CW}_\text{min}}{2}.
\end{equation}

On the other hand, if the two users are hidden terminals to each other, the analysis of the optimal CCT is more involved.

\begin{prop}\label{lem:optimal_CSMA_DRL_hidden}
Consider a two-user heterogeneous network, where one user operates with the basic mode CSMA/CA, the other user operates with DRLMA, and the two users are hidden terminals to each other.
The minimum achievable CCT of the network is given by
\begin{eqnarray}\label{eq:V-11}
\Psi^*_\text{DRLMA-hidden} =
\sum_{I=0}^{+\infty} &&\hspace{-0.6cm} \left[ p_{s,I}\prod_{i=0}^{I-1}(1-p_{s,i}) \right] \times \\
&&\hspace{-3cm} \left[ \sum_{i=0}^{I} (\ell_\text{difs}+\ell_\text{tran}+\mathbb{E}(\widetilde{\lambda}_i) )+
\ell_\text{difs}+\ell_\text{tran}+\mathbb{E}(\widetilde{\lambda}_{\rho+1}^{\prime}) \right]. \nonumber 
\end{eqnarray}
where 
$$p_{s,i}= \begin{cases}
\frac{\text{CW}_i-(\ell_\text{tran}-\ell_\text{difs})}{\text{CW}_i-1},&i\ge \gamma,\\
0,&i<\gamma.
\end{cases}$$
and $\gamma=\left\lceil \log_2\left( \frac{\ell_\text{tran}-\ell_\text{difs}}{\text{CW}_\text{min}} \right) \right\rceil$, and
\begin{equation}\label{eq:V-13}
\widetilde{\lambda}_i \sim \begin{cases}
    U(1,\text{CW}_i), & i<\gamma, \\
    U(1,\ell_\text{tran}-\ell_\text{difs}-T_\text{slot}), & i\ge \gamma. \\
\end{cases}
\end{equation}
\end{prop}

\begin{NewProof}
Suppose user A operates with the basic mode CSMA/CA and user B operates with DRLMA. Consider the following transmission policy of user B:
\begin{enumerate}[leftmargin=0.5cm]
\item User B takes the initiative to transmit packets, regardless of user A's states, until it achieves a successful transmission.
\item Once a packet is transmitted successfully, user B stays silent and waits until an ACK broadcasted by the AP, indicating A's successful transmission. Go to step 1).
\end{enumerate}

Unlike the optimal policy in the proof of Proposition \ref{lem:optimal_CSMA_DRL}, here, user B's transmission in step 1) is very likely to collide with user A's transmissions (since they are hidden terminals). User B might need to make multiple transmission attempts to achieve a successful transmission.

It can be proven that the above policy is optimal and there exists no transmission policy of user B that leads to a smaller CCT. First, for User A, each successful transmission resets its contention window size and counter. As a result, there must be at least one successful transmission from User B between two consecutive successful transmissions of user A; otherwise, the cycle time of user A increases. Nevertheless, the randomly generated counter value of user A might be too small, preventing user B from successfully transmitting a packet. Therefore, user B should persistently attempt transmissions, intentionally colliding with user A's transmission. This ultimately leads to user A generating a sufficiently large counter value, enabling user B to achieve a successful transmission. It's important to note that, at this point, user B can and should transmit only once, as multiple transmissions would elongate the cycle times of both A and B.

\begin{figure}[t]
\centering
\includegraphics[width=0.9\columnwidth]{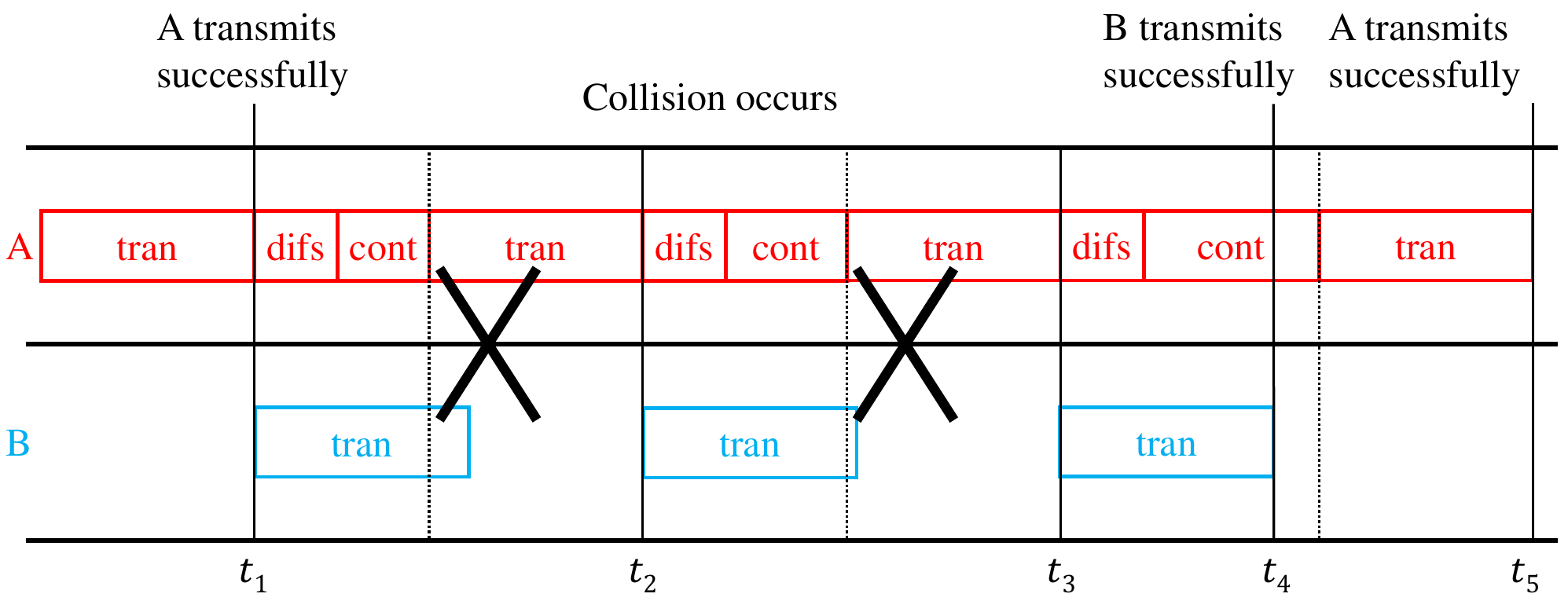}
\caption{The optimal transmission policy of the DRLMA user in Proposition \ref{lem:optimal_CSMA_DRL_hidden}.}
\label{fig:CSMA_DRL_hidden_optimal}
\end{figure}

Now, we derive the corresponding optimal CCT. Consider the cycle time of user A.
As shown in Fig.~\ref{fig:CSMA_DRL_hidden_optimal}, in each cycle time of A, there is one successful transmission of B.
When the backoff counter of A is smaller than $\ell_\text{difs} + \ell_\text{tran}$, a collision occurs.
In backoff stage $i~(0\le i \le \beta)$, the probability that user B can have a successful transmission is
\begin{equation}\label{eq:V-12}
p_{s,i}= \Pr(\lambda_i+\ell_\text{difs} \ge \ell_\text{tran}) = \begin{cases}
\frac{\text{CW}_i-(\ell_\text{tran}-\ell_\text{difs})}{\text{CW}_i-1},&i\ge \gamma,\\
0,&i<\gamma,
\end{cases}
\end{equation}
where $\gamma$ is the minimum value such that $2^\gamma\text{CW}_\text{min}+\ell_\text{difs} \ge \ell_\text{tran}$:
\begin{equation}
\gamma=\left\lceil \log_2\left( \frac{\ell_\text{tran}-\ell_\text{difs}}{\text{CW}_\text{min}} \right) \right\rceil
\end{equation}

When user A is in stage $i$, a collision occurs with probability $1-p_{s,i}$, and the corresponding time consumed is $\ell_\text{difs}+\ell_\text{tran}+\widetilde{\lambda}_i$, where $\widetilde{\lambda}_i$ is the random backoff counter under the condition that it is smaller than $(\ell_\text{tran}-\ell_\text{difs})$, and we have \eqref{eq:V-13}.

Then, the cycle time of A can be expressed as
\begin{equation}\label{eq:V-14}
\Gamma_A (I)=\sum_{i=0}^{\rho} (\ell_\text{difs}+\ell_\text{tran}+\widetilde{\lambda}_i)+
\ell_\text{difs}+\ell_\text{tran}+\widetilde{\lambda}_{\rho+1}^{\prime},
\end{equation}
where $\widetilde{\lambda}_{\rho+1}^{\prime}$ is the random backoff counter in the $(\rho+1)$-th stage under the condition that it is larger than $(\ell_\text{tran}-\ell_\text{difs})$  such that B can transmit a packet successfully, and $\widetilde{\lambda}_{\rho+1}^{\prime} \sim U(\ell_\text{tran}-\ell_\text{difs}, \text{CW}_{\rho+1})$;
$\rho$ is the number of collisions, and its distribution is given by
\begin{equation}\label{eq:V-15}
\Pr(\rho=I)=p_{s,I}\prod_{i=0}^{I-1}(1-p_{s,i}),~
I=0,1,2,\cdots
\end{equation}

The average cycle times of users A and B are
\begin{eqnarray}\label{eq:V-16}
\mathbb{E}(\Gamma_A)=\mathbb{E}(\Gamma_B)=
\sum_{I=0}^{+\infty} \mathbb{E}[\Gamma_A (I)] \cdot \Pr(\rho=I).
\end{eqnarray}
Substituting \eqref{eq:V-14} and \eqref{eq:V-15} into \eqref{eq:V-16} gives us \eqref{eq:V-11}.
\end{NewProof}

To conclude this section, let us take a step back and compare the advantages of CCT with the average number of inter-transmissions, which is another easily calculable short-term fairness measure, in the context of this two-user heterogeneous network. Suppose the intelligent node's objective is to optimize the average number of inter-transmissions, as opposed to optimizing CCT for the network.
In this scenario, the optimal transmission pattern we have arrived at may be 'AABBAABBAA...'. This is because `AABBAABBAA...' and `ABABABABAB...' are equivalent as far as the average number of inter-transmissions is concerned.
However, `ABABABABAB...', which is obtained from optimizing CCT, is the one that is more evidently fair in the short term.

\section{Simulation results}\label{sec:VI}
This section presents analytical and simulation results to evaluate the CCT performance of various MAC protocols and scenarios analyzed in this paper, including homogeneous networks operated with slotted Aloha or CSMA/CA, and a heterogeneous network with a CSMA/CA user and a DRLMA user. 
The parameter settings for the simulations are summarized in Table~\ref{tab:paras}.

\begin{table}
\centering
\caption{Parameter settings.}
\begin{tabular}{cccc}
\toprule
\multirow{2}{*}{Protocols}     & \multicolumn{3}{c}{Parameters}  \\
                               & Descriptions & Symbols & Values \\
\midrule
round-robin TDMA  & slot/packet duration  & $T_\text{slot}$   & $0.6 \sim 6$ms  \\
  \midrule
\multirow{2}{*}{slotted Aloha} & trans. Pr. of each user & $p$      & $0 \sim 1$           \\
                               & slot/packet duration                       & $T_\text{slot}$   & $0.6 \sim 6$ms  \\
  \midrule
\multirow{8}{*}{\begin{tabular}[c]{@{}c@{}} {CSMA/CA} \\ \end{tabular}}
                        & duration of a slot     & $T_\text{slot}$        & $20\mu$s  \\
                        & minimum CW size        & $\text{CW}_\text{min}$ & $32\cdot T_\text{slot}$             \\
                        & maximum CW size        & $\text{CW}_\text{max}$ & $1024\cdot T_\text{slot}$           \\
                        & duration of DIFS       & $\ell_\text{difs}$     & $80\mu$s           \\
                        & duration of ACK        & $\ell_\text{ack}$      & $20\mu$s           \\
                        & duration of RTS        & $\ell_\text{rts}$      & $20\mu$s           \\
                        & duration of CTS        & $\ell_\text{cts}$      & $20\mu$s           \\
                        & duration of a packet   & $\ell_\text{pkt}$      & $0.6 \sim 6$ms \\
\bottomrule
\end{tabular}
\label{tab:paras}
\end{table}

\subsection{A homogeneous network with slotted Aloha}\label{sec:VI-A}
\begin{figure}[t]
\centering
\includegraphics[width=0.72\columnwidth]{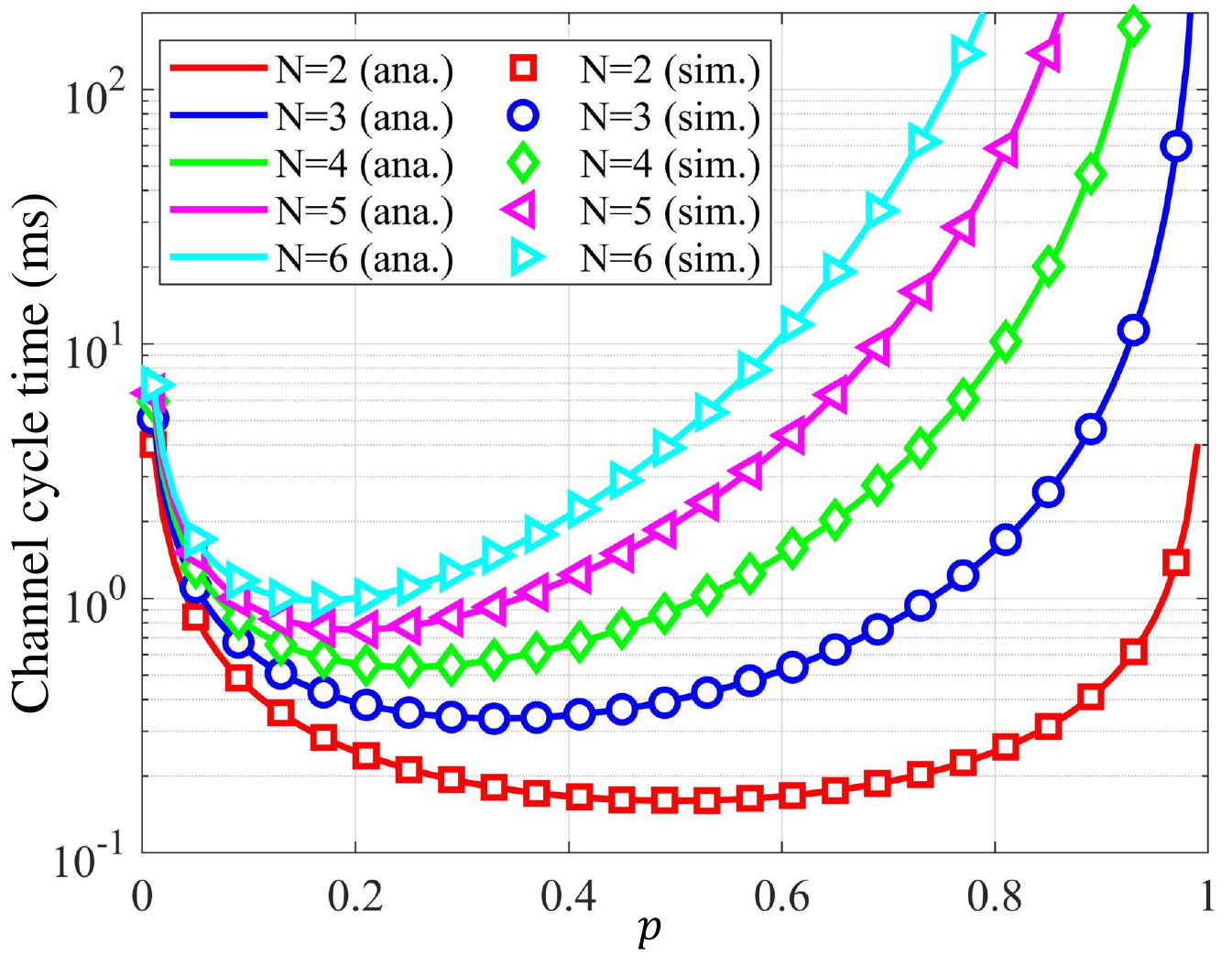}
\caption{CCT of a homogeneous network with slotted Aloha, where $T_\text{slot}=20\mu$s.}
\label{fig:result_ALOHA}
\end{figure}

We first evaluate the CCT performance of a homogeneous network operated with slotted Aloha.
To validate our derivations, we compare the analytical and simulation results of CCT versus the user transmission probability $p$ under various user numbers $N$ in Fig.~\ref{fig:result_ALOHA}.
We have three main observations:
\begin{itemize}[leftmargin=0.45cm]
    \item The simulation results are aligned with our analytical results very well for all cases.
    \item For any user number $N$, CCT decreases first and then increases, as we increase $p$ from 0 to 1.
    This observation matches our intuition: when $p$ is small, there are not many packets in the channel, thus the packets can be transmitted more frequently as $p$ increases, leading to a smaller CCT. 
    On the other hand, as $p$ becomes larger and larger, the collision probability increases, resulting in a waste of channel resources and an increase in the CCT.
    \item If CCT is used as the design principle to optimize slotted Aloha, the user transmission probability should be set to $p^*=1/N$, which corroborates our analytical results in Section~\ref{sec:III-B}.
\end{itemize}

\begin{figure}[t]
\centering
\includegraphics[width=0.72\columnwidth]{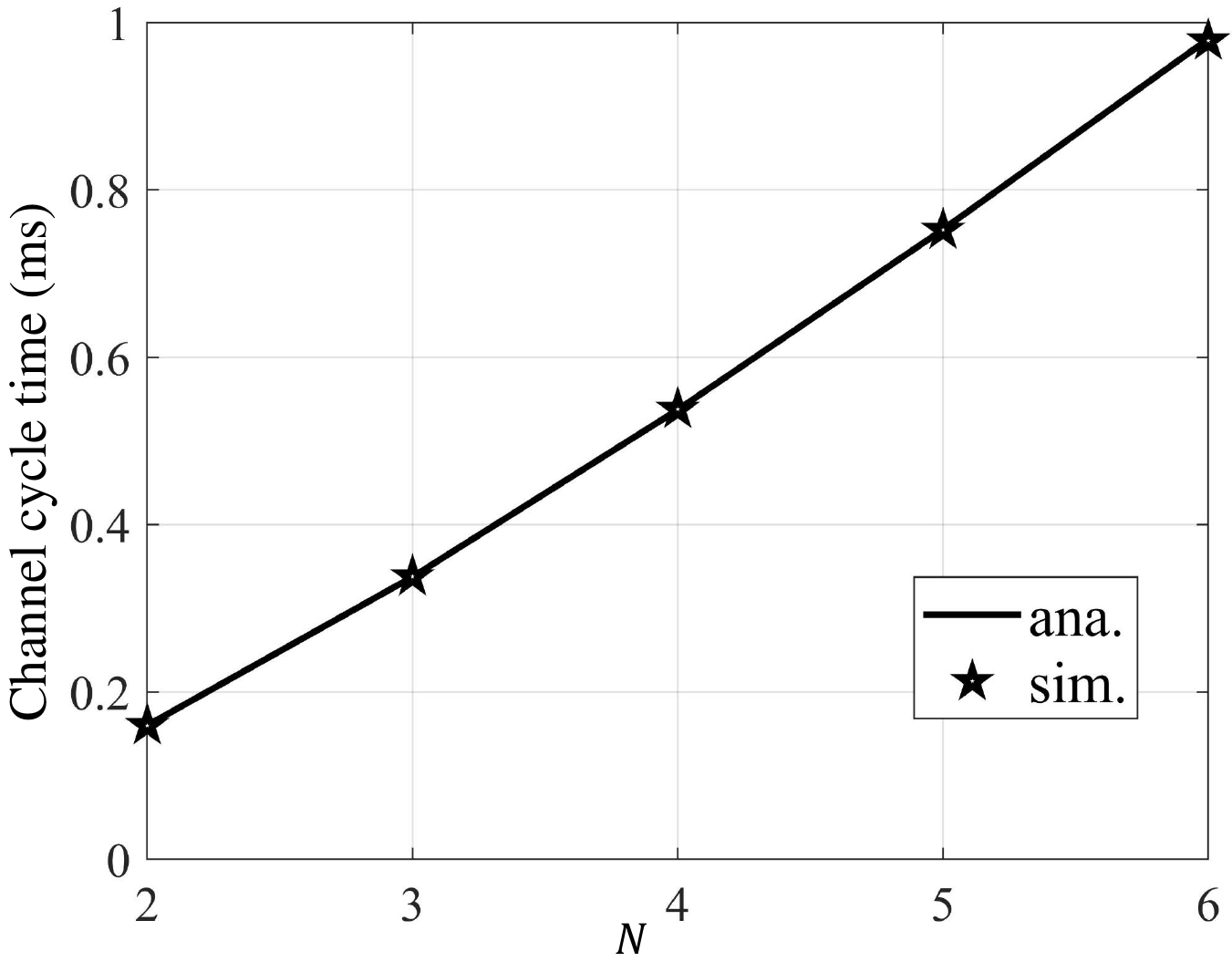}
\caption{The optimal CCT of a homogeneous network operated with slotted Aloha, where $T_\text{slot}=20\mu$s.}
\label{fig:result_ALOHA_optimal}
\end{figure}

In Fig.~\ref{fig:result_ALOHA_optimal}, we present the optimal CCT of the homogeneous network with slotted Aloha under different user numbers.
As shown, the optimal CCT value increases as the number of users increases, underscoring the relationship between CCT and access delay.

\subsection{A homogeneous network with CSMA/CA}\label{sec:VI-B}
This section evaluates the CCT of CSMA/CA in a homogeneous network.
We shall first consider the two-user case to validate our analytical results, and then extend the simulations to the case with more users.

\begin{figure}[t]
\centering
\includegraphics[width=0.75\columnwidth]{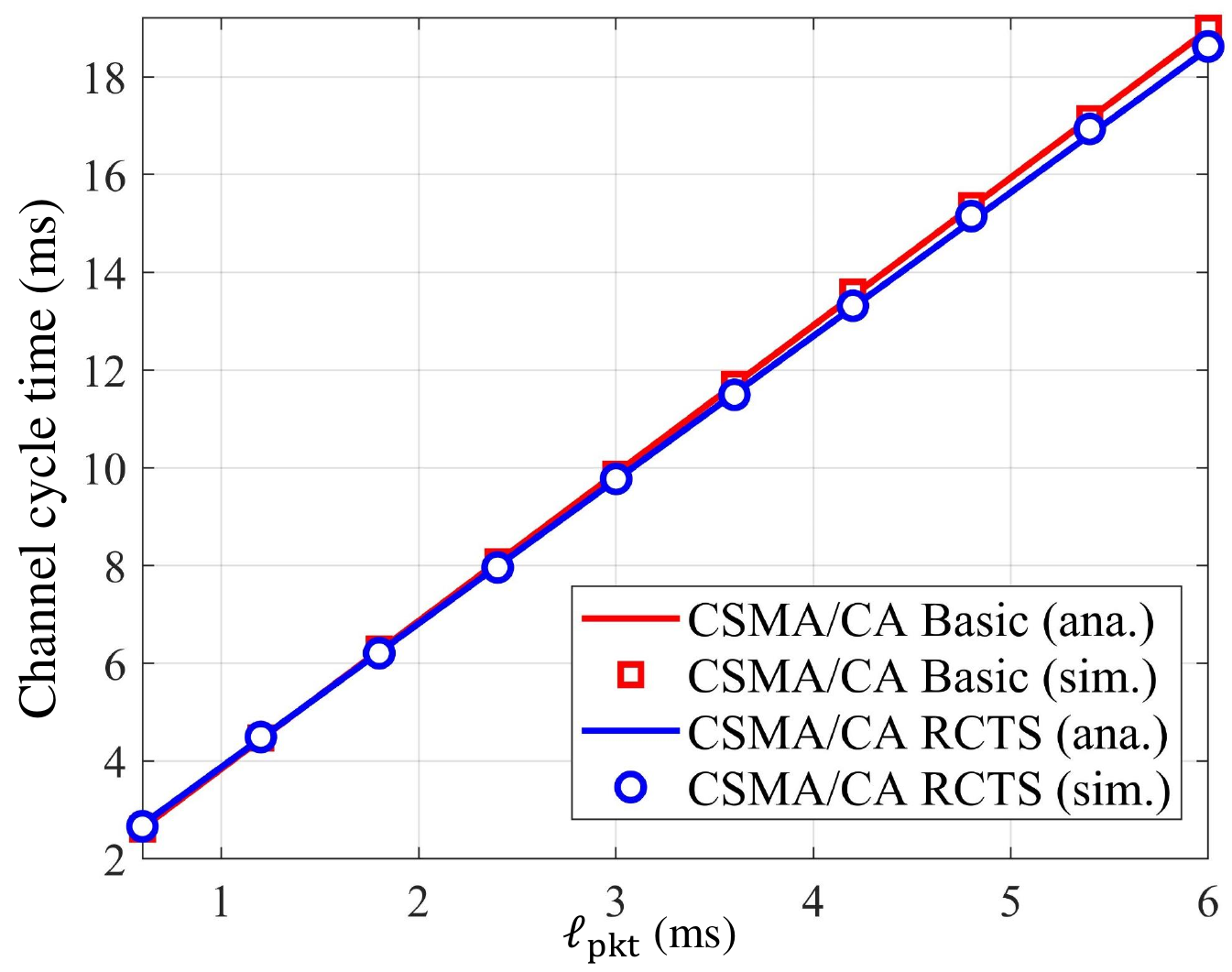}
\caption{CCT of CSMA/CA in the basic and RTS/CTS modes in a two-user homogeneous network, where the two users are audible to each other.}
\label{fig:result_CSMA}
\end{figure}

Fig.~\ref{fig:result_CSMA} compares the CCT of CSMA/CA in the RTS/CTS mode and the basic mode in a two-user homogeneous network, in which the two users are audible to each other (i.e., they are not hidden terminals).
As can be seen, the analytical results of CSMA/CA match the simulation results very well, validating our analysis.
It also shows that the RTS/CTS mode exhibits a larger (smaller) channel cycle time than the basic mode when $\ell_\text{pkt}$ is small (large).
The inflection point occurs when
$$\ell_\text{pkt}= \frac{2-p_c}{p_c}\ell_\text{rcts} \approx 1.44 \text{ms}.$$
This is easy to understand since the RTS/CTS mode introduces additional handshake overhead before transmission to reduce the collision probability.
When the packet duration is short, the reduced collision probability by RTS/CTS does not outweigh the incurred time costs .

\begin{figure}[t]
\centering
\includegraphics[width=0.75\columnwidth]{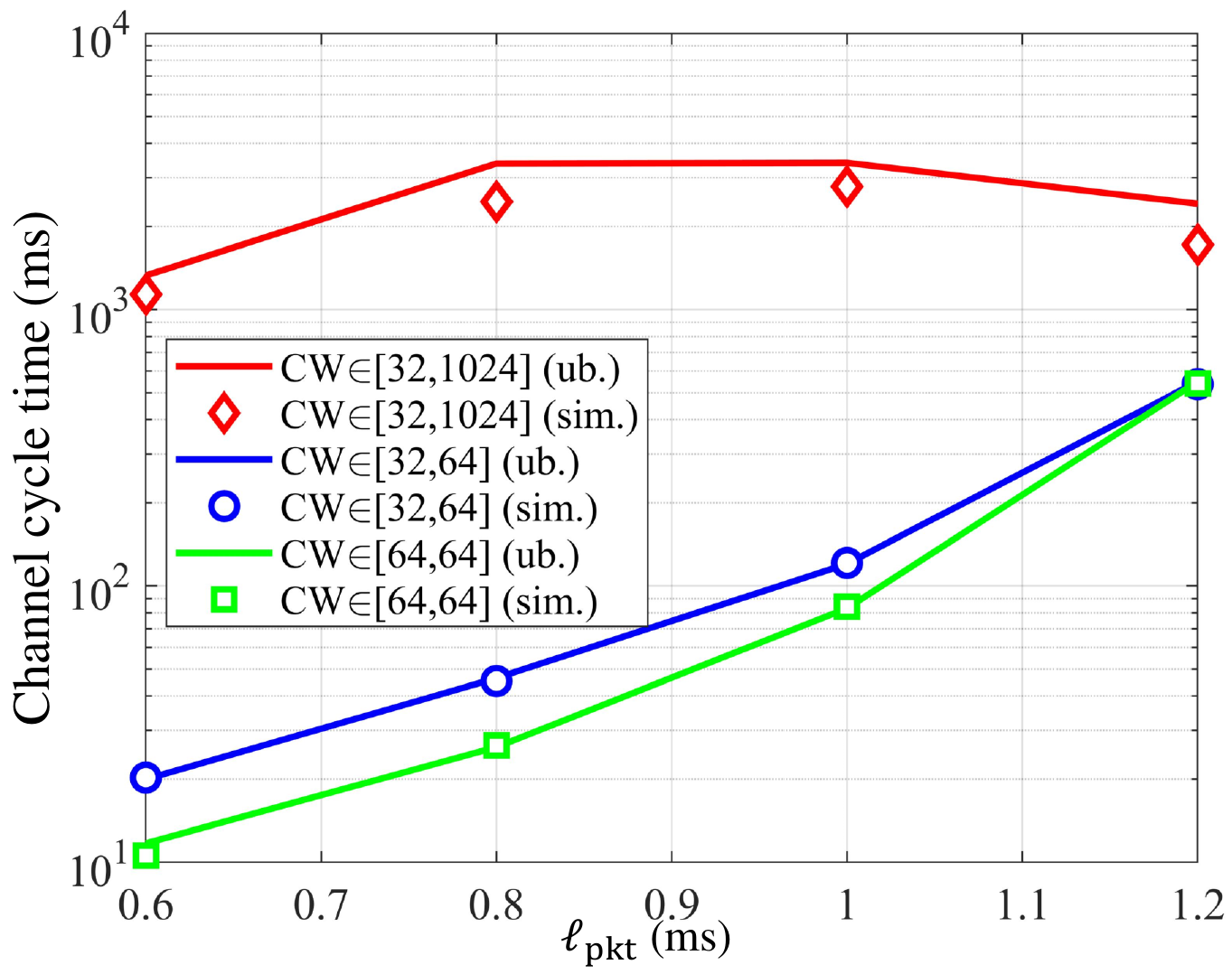}
\caption{CCT of CSMA/CA in the basic and RTS/CTS modes in a two-user homogeneous network, where the two users are hidden terminals to each other.}
\label{fig:result_CSMA_hidden}
\end{figure}

When the two users are hidden terminals to each other, on the other hand, Fig.~\ref{fig:result_CSMA_hidden} compares the CCT of CSMA/CA under the basic and RTS/CTS modes.
We have the following observations:
\begin{itemize}[leftmargin=0.45cm]
\item We have proven in Theorem~\ref{thm:CW_bounds_CSMA_hidden} that our upper bound is tight when $\text{CW}_\text{max}=\text{CW}_\text{min}$ or $\text{CW}_\text{max}=2\text{CW}_\text{min}$. This is confirmed in Fig.~\ref{fig:result_CSMA_hidden} when $\text{CW}_\text{min}=\{32,64\}\cdot T_\text{slot}$ and $\text{CW}_\text{max}=64\cdot T_\text{slot}$.
\item If we fix $\text{CW}_\text{min}$, a larger $\text{CW}_\text{max}$ results in a larger CCT.
This is because the two users are hidden terminals. Frequent collisions will result in one user getting the largest contention window, while the other user gets the smallest contention window, leading to the phenomenon of CCT increasing as $\text{CW}_\text{max}$ increases.
\item The increase in $\text{CW}_\text{min}$ leads to a smaller CCT. This can be understood in that a large $\text{CW}_\text{min}$ results in a lower collision probability in the presence of hidden terminals. Therefore, CCT decreases.
\end{itemize}

\begin{figure}
    \centering
    \begin{subfigure}[t]{0.75\linewidth}
        \centering
        \includegraphics[width=\linewidth]{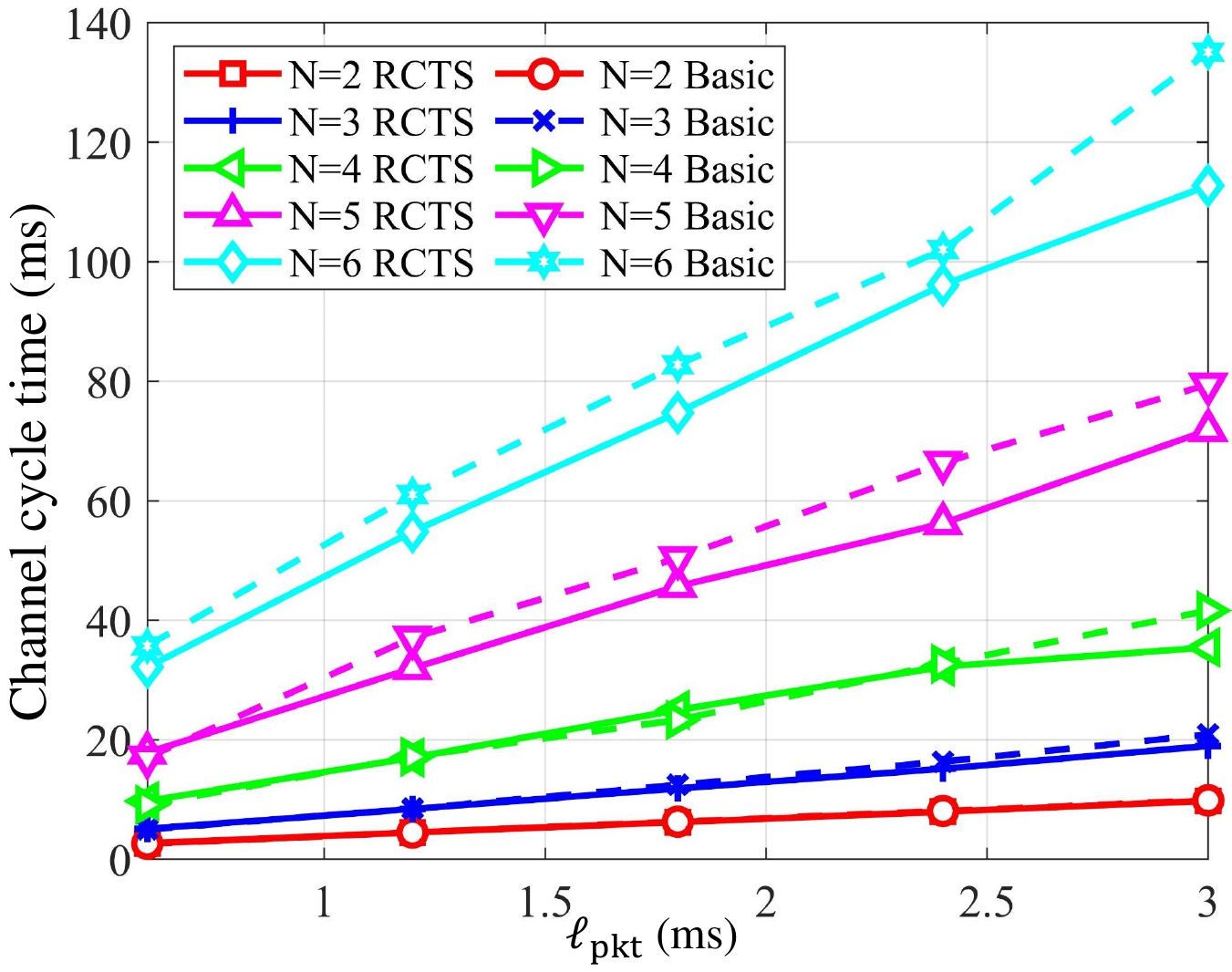}
        \caption{CCT versus $\ell_\text{pkt}$.}
    \end{subfigure}
    \qquad
    \begin{subfigure}[t]{0.75\linewidth}
        \centering
        \includegraphics[width=\linewidth]{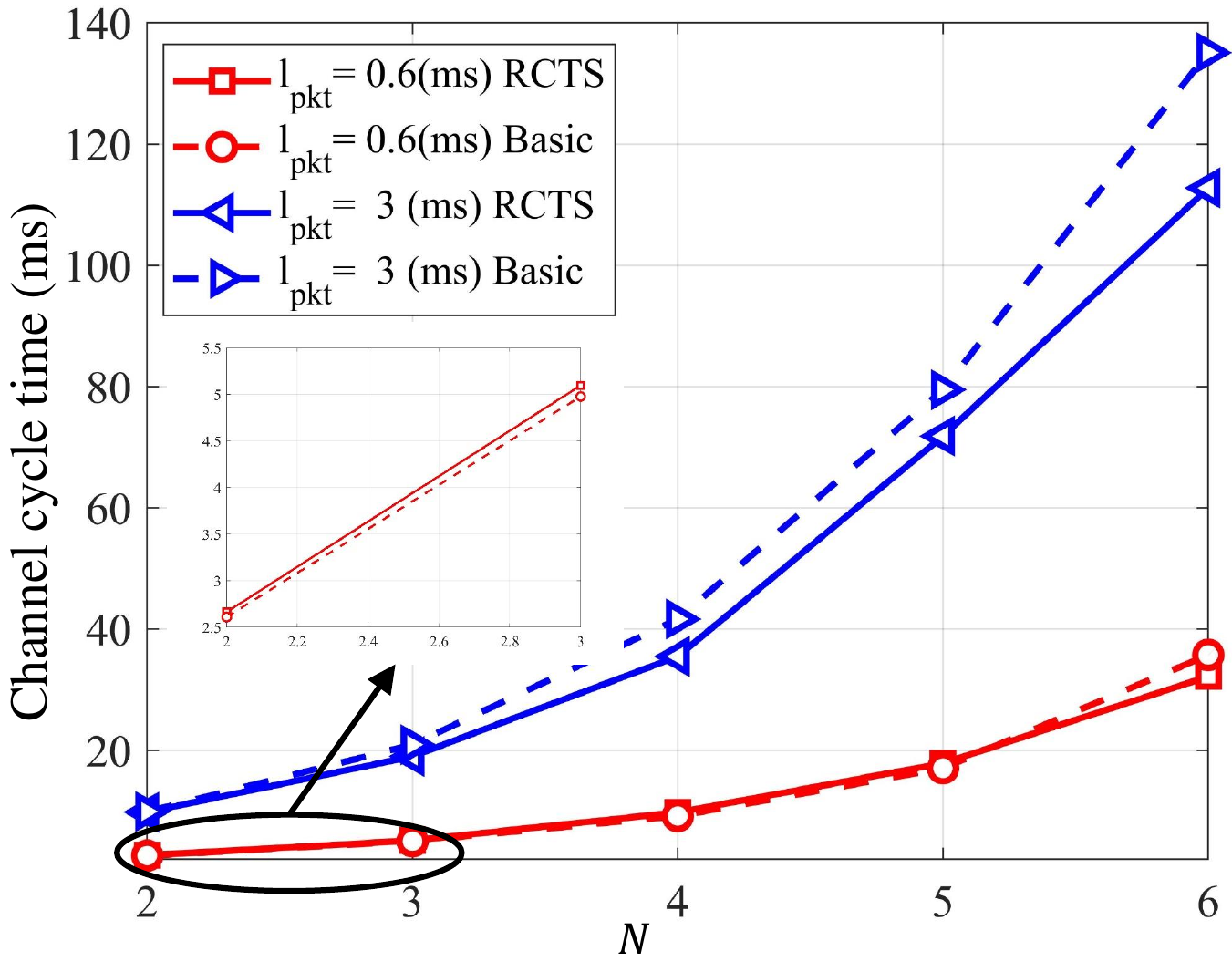}
        \caption{CCT versus $N$.}
    \end{subfigure}\\
    \caption{CCT of CSMA/CA in the basic and RTS/CTS modes in a homogeneous network with more than two users. There are no hidden terminals in the network.}    
    \label{fig:result_CSMA_moreuser}
\end{figure}

Next, we consider homogeneous networks with more than two users and evaluate the CCT of CSMA/CA. We assume there are no hidden terminals in the network.
The simulation results are given in Fig.~\ref{fig:result_CSMA_moreuser}, where (a) presents the CCT versus $\ell_\text{pkt}$ and (b) presents the CCT versus $N$.

As can be seen from Fig.~\ref{fig:result_CSMA_moreuser}(a),
\begin{itemize}[leftmargin=0.45cm]
    \item In both the basic and RTS/CTS modes of CSMA/CA, CCT increases almost linearly in the packet length $\ell_\text{pkt}$.
    \item The CCT of CSMA/CA in the RTS/CTS mode is smaller than that in the basic mode when $\ell_\text{pkt}$ is large; while they can be larger when $\ell_\text{pkt}$ and $N$ are small (as shown in Fig.~\ref{fig:result_CSMA}).
    Therefore, for large $\ell_\text{pkt}$, the RTS/CTS mode is superior to the basic mode as far as the short-term fairness is concerned.
\end{itemize}

Fig.~\ref{fig:result_CSMA_moreuser}(b) shows that the CCT increases in the number of users $N$. In particular, the gains of the RTS/CTS mode over the basic mode is more pronounced in the case of larger $N$.
This is intuitive as more users means larger collision probabilities without RTS/CTS.

\subsection{The two-user heterogeneous network}\label{sec:VI-C}
This section evaluates the CCT of a two-user heterogeneous network with a CSMA/CA user and a DRLMA user and explores the minimum achievable CCT in such a network.

\begin{figure}[t]
\centering
\includegraphics[width=0.75\columnwidth]{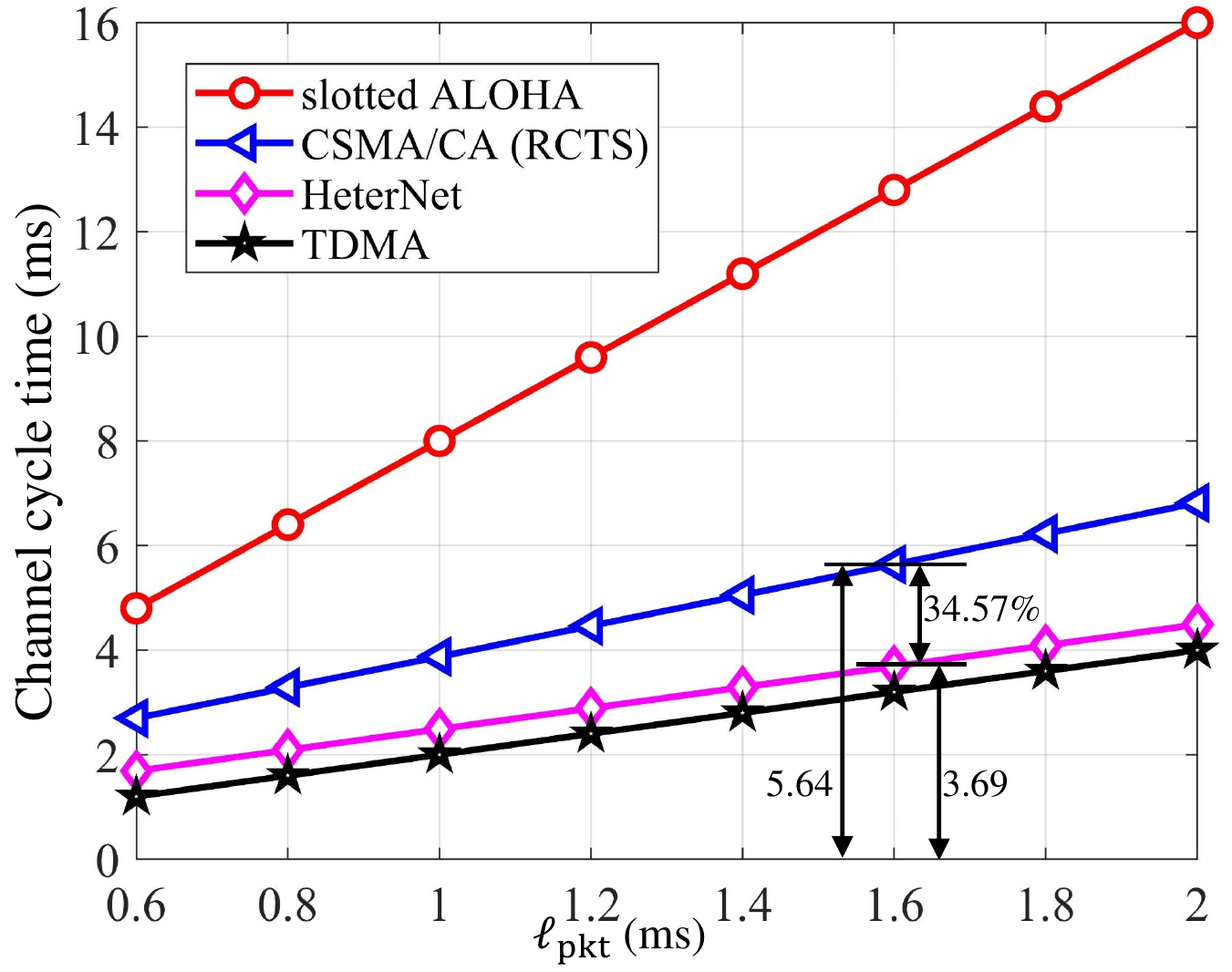}
\caption{CCT of the two-user heterogeneous network benchmarked against the homogeneous networks operated with slotted Aloha, CSMA/CA, and TDMA. The CSMA/CA user in the heterogeneous network is operated in the RTS/CTS mode.}
\label{fig:result_DRLMA_compare_all}
\end{figure}

First, we consider the case that the CSMA/CA user in
the heterogeneous network is operated in the RTS/CTS mode. Fig.~\ref{fig:result_DRLMA_compare_all} shows the minimum achievable CCT benchmarked against the homogeneous networks operated with slot-
ted Aloha, CSMA/CA, and round-robin TDMA (which is known to be the short-term fairest).

As can be seen,
\begin{itemize}[leftmargin=0.45cm]
    \item CSMA/CA is a short-term fairer protocol than slotted Aloha.\footnote{This is aligned with the results of prior works that use other short-term fairness measures.} With the DRLMA user, the heterogeneous network (abbreviated as ``HeterNet'') can further reduce the CCT of CSMA/CA by $34.57\%$.
    \item The CCT of CSMA/CA and HeterNet exhibit a relatively fixed proportional gap with that of round-robin TDMA.
    The reason behind this is that, unlike slotted Aloha, the negotiation overhead of CSMA/CA and HeterNet is relatively constant to $\ell_\text{pkt}$, which is revealed by \eqref{eq:IV-7} and \eqref{eq:V-17}. Therefore, the slopes of CCT of CSMA/CA, HeterNet and round-robin TDMA w.r.t. $\ell_\text{pkt}$ are on an equal footing.
\end{itemize}

\begin{figure}[t]
\centering
\includegraphics[width=0.75\columnwidth]{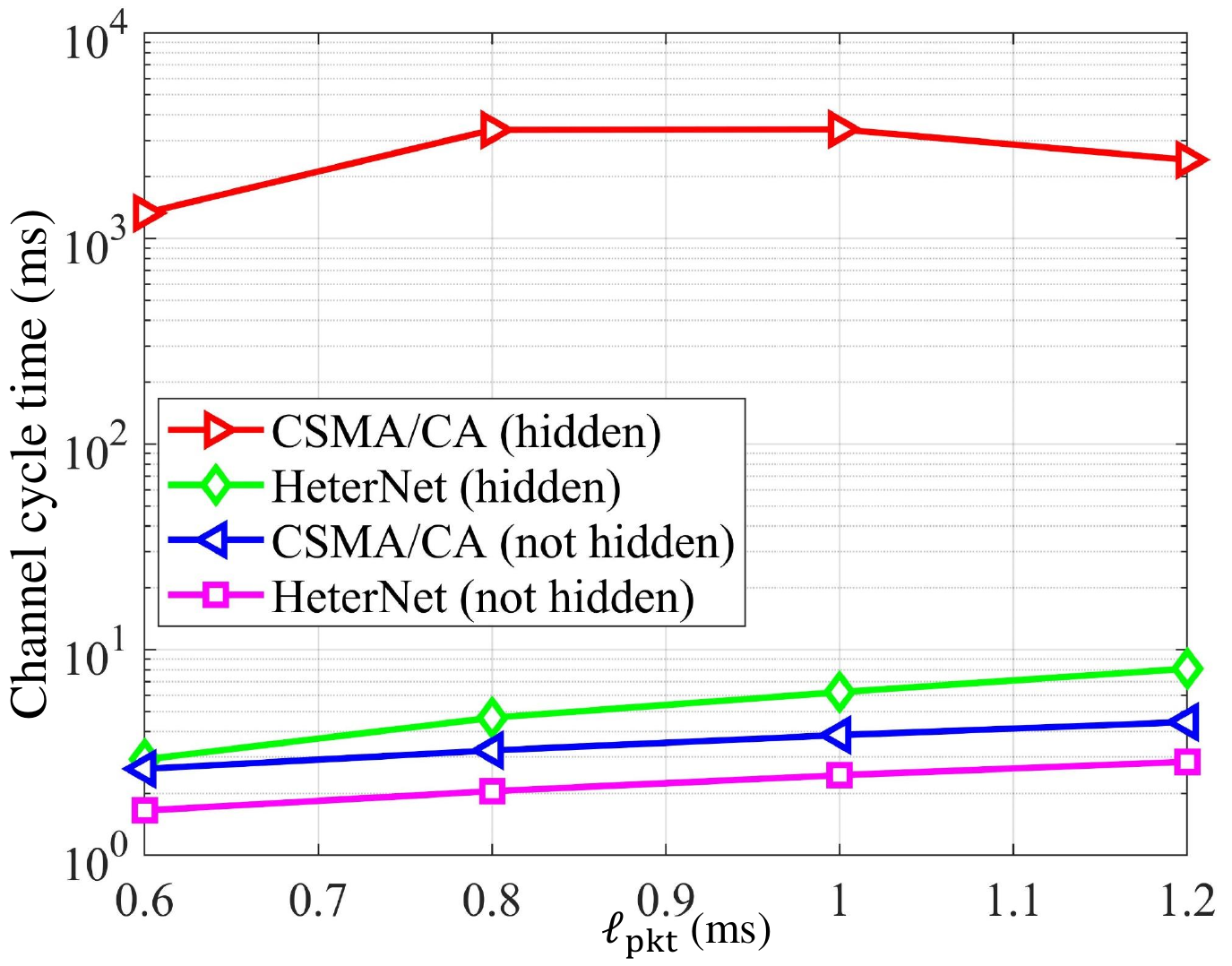}
\caption{CCT of the two-user heterogeneous network benchmarked against the homogeneous networks operated with slotted Aloha, CSMA/CA, and TDMA. The CSMA/CA user in the heterogeneous network is operated in the basic mode.}
\label{fig:result_DRLMA_hidden}
\end{figure}

Then, we consider the case that the CSMA/CA user in the heterogeneous network is operated in the basic mode and compare the CCT of CSMA/CA and HeterNet.
The simulation results are shown in Fig.~\ref{fig:result_DRLMA_hidden}.
A first straightforward observation is that the CCT is larger if the two users are hidden terminals than the case where the two users are audible to each other.
However, in the HeterNet, the performance degradation caused by hidden terminals can be significantly alleviated, thanks to the presence of intelligent user.

\section{Conclusion}\label{sec:VII}
Short-term fairness plays a crucial role in real-time applications. Conventional methods primarily focused on successful transmissions and employed a set of values or distributions to assess short-term fairness.
In this paper, we introduced and thoroughly explored the concept of channel cycle time (CCT) as a metric for measuring short-term fairness in multiple-access networks. This metric, which characterizes the average duration between two successful transmissions of a user, during which all other users have successfully accessed the channel at least once, offers a fresh perspective on evaluating the transient behavior of MAC protocols. 
Moreover, CCT's emphasis on users' delay provides a more comprehensive view of short-term fairness, aligning with the evolving needs of modern networks. 

The demonstrated effectiveness of CCT through the comparison of two classical MAC protocols, slotted Aloha and CSMA/CA, underscores its practical utility. The analytical derivation of closed-form CCT values reveals that CSMA/CA outperforms slotted Aloha in terms of short-term fairness, validating the metric's discriminatory power. Beyond its role as a metric, CCT can be used as a guiding principle in MAC protocol design. By strategically optimizing CCT during the development process, we devised MAC protocols for a two-user heterogeneous network that excel in short-term fairness.

\appendices
\section{Proof of Lemma~\ref{lem:part1} and Lemma~\ref{lem:part2}}\label{sec:AppA}
\begin{figure}[t]
\centering
\includegraphics[width=0.9\columnwidth]{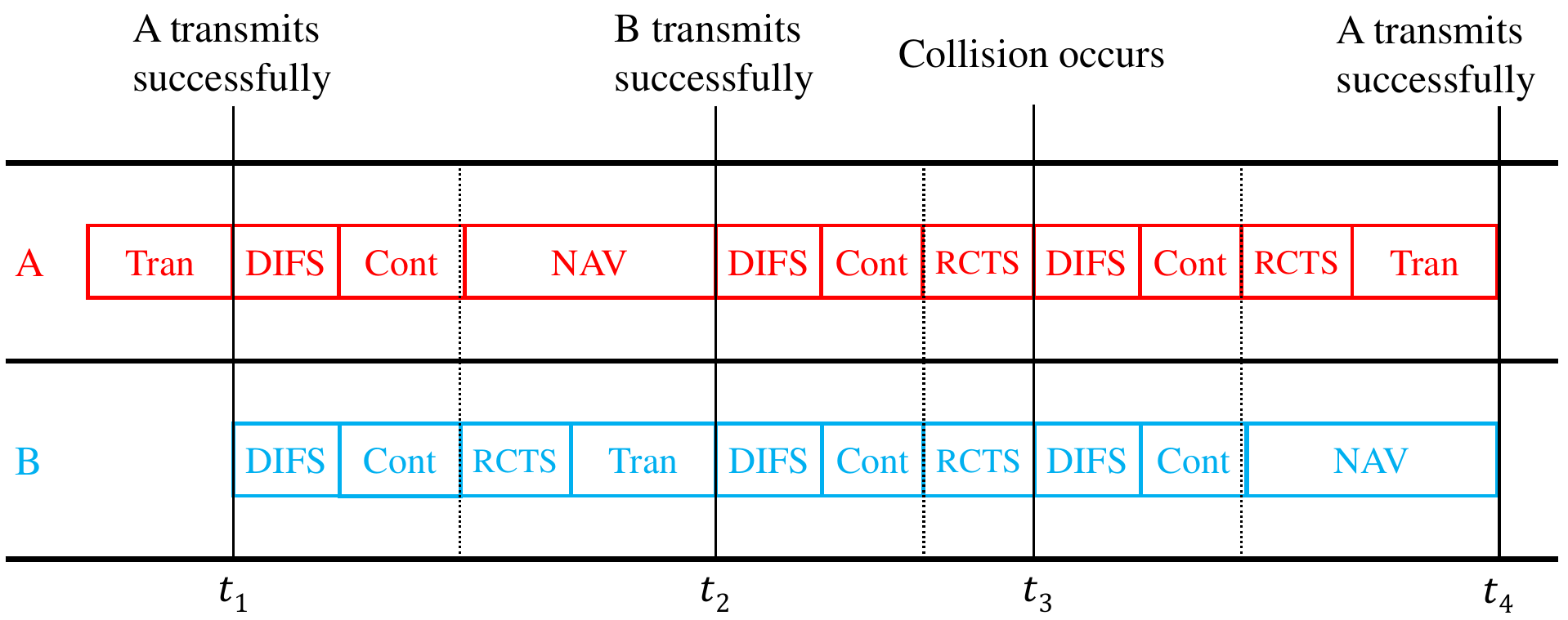}
\caption{CSMA/CA: states of users in Part 1.}
\label{fig:actions_part1_rcts}
\end{figure}


We first prove Lemma \ref{lem:part1}.
In Part 1, user A can experience five different states: DIFS, Cont (contention), NAV, RCTS (RTS and CTS), and Tran (transmission). For a more intuitive understanding, Fig.~\ref{fig:actions_part1_rcts} gives a concrete example to show the states of users in Part 1, where there is one successful transmission of user B (at $t_2$), one collision (at $t_3$), and two successful transmissions of A (at $t_1$ and $t_4$).
The duration of Part 1 is the sum of the duration of the five states. Specifically,
\begin{itemize}[leftmargin=0.4cm]
    \item For every successful transmission of user B, user A experiences a DIFS state and a NAV state, which takes $(\ell_\text{difs} + \ell_\text{nav})$. In total, it takes $n_B(\ell_\text{difs} + \ell_\text{nav})$ for user A.
    \item The only successful transmission of user A takes a state of Tran, which takes $\ell_\text{tran}$.
    \item For each collision in Part 1, user A experiences a DIFS state and an RCTS state. The random backoff counter $\lambda_i$ is selected according to a doubled contention window after each collision. Overall, assuming that there are $\rho_1$ collisions, it takes  $$\sum_{i=0}^{\rho_1} (\ell_\text{difs} + \ell_\text{rcts} + \lambda_i).$$
\end{itemize}

Summing up the time consumed above gives us $T_\text{part1}$.

Lemma \ref{lem:part2} can be proven in a similar manner.
In Part 2, user A has $n_A^\prime$ successful transmissions and user B has no successful transmission.
Thus, user A will only experience DIFS, Cont, RCTS, and Tran.

For the $j$-th ($j=0, 1, \cdots, n_A^\prime$) successful transmission of user A, we assume there are $\rho_2^{(j)}$ collisions. The time consumed by transmitting the $j$-th packet is then
$$\ell_\text{tran} + \sum_{i=0}^{\rho_2^{(j)}} \left( \ell_\text{difs} + \ell_\text{rcts} + \lambda_i^{(j)} \right).$$
Eq. \eqref{eq:IV-5} follows by summing up the time consumed by $n_A^\prime$ successful transmissions of user A.

\section{Proof of Theorem~\ref{prop:CCT_CSMA_basic}}\label{sec:AppB}
To characterize the channel cycle time of CSMA/CA, we first derive the average cycle time of user A:
$$\mathbb{E}( \Gamma_A )= 
    \mathbb{E}( T_\text{part1} ) + \mathbb{E}( T_\text{part2} ).$$

From Lemmas \ref{lem:nAB} and \ref{lem:part1}, we have
\begin{equation}\label{eq:AppB-1}
    \mathbb{E} (T_\text{part1}) \!=\! \frac{(\ell_\text{difs} \!+\! \ell_\text{nav}) \!\cdot\! \mathbb{E}(N_I)}{1-P_{N_I,0}} \!+\! \ell_\text{tran} \!+\! \frac{\ell_\text{difs} \!+\! \ell_\text{rcts}}{1-p_c} \!+\! 
    \mathbb{E} \! \left(\sum_{i=0}^{\rho_1} \lambda_i \right), 
\end{equation}
where the average number of inter-transmissions $\mathbb{E}(N_I)=1$ \cite{berger2005short}, $p_c$ is the collision probability of CSMA/CA, and 
$$\mathbb{E}(\rho_1)= \sum_{i=0}^{\infty} i \cdot p_c^i (1-p_c) =\frac{p_c}{1-p_c}.$$

Note that the state transition of CSMA/CA can be formulated as a Markov Chain, based on which the collision probability $p_c$ can be derived. The derivation is the same as that in \cite{840210}, and we can obtain the recursive equation \eqref{eq:IV-10} with a unique solution.

From Lemmas \ref{lem:nAB} and \ref{lem:part2}, we have
\begin{equation}\label{eq:AppB-2}
    \mathbb{E} (T_\text{part2}) = \frac{P_{N_I,0}}{1-P_{N_I,0}} \!\! \left[ \frac{\ell_\text{difs} + \ell_\text{rcts}}{1-p_c} + \ell_\text{tran} + 
    \mathbb{E}\left(\sum_{i=0}^{\rho_2} \lambda_i \right) \right].
\end{equation}
Note that $\mathbb{E}(\rho_2^{(j)})=\frac{p_c}{1-p_c}$, $j=0,1,\cdots,n_A^\prime$.
Since $\{\rho_2^{(j)}\}$ are i.i.d. random variables, we denote by $\rho_2$ the random variable representing the number of collisions that occurred in Part 2 for each successful transmission of user A.

Then, the average cycle time of user A can be written as
\begin{eqnarray}\label{eq:AppB-3}
    \mathbb{E}( \Gamma_A ) &&\hspace{-0.7cm} = 
    \mathbb{E}( T_\text{part1} ) + \mathbb{E}( T_\text{part2} ) \\
    &&\hspace{-1.0cm} = \frac{1}{1\!-\!P_{N_I,0}} \! \left[ \ell_\text{difs} \!+\! \ell_\text{nav} \!+\! \ell_\text{tran} \!+\! \frac{\ell_\text{difs} \!+\! \ell_\text{rcts}}{1\!-\!p_c} \!+\! \mu \right],  \nonumber    
\end{eqnarray}
where
\begin{equation*}
\mu \triangleq
 (1-P_{N_I,0}) \mathbb{E} \left(\sum_{i=0}^{\rho_1} \lambda_i \right) + P_{N_I,0} \mathbb{E}  \left(\sum_{i=0}^{\rho_2} \lambda_i \right).
\end{equation*}

The exact form of $\mu$ can be derived as follows:
\begin{eqnarray}\label{eq:AppB-4}
\mu = &&\hspace{-0.6cm} \sum_{i=0}^{\beta} \! \left[ p_c^i(1-p_c) \cdot \sum_{j=0}^i \mathbb{E}(\lambda_j) \right] \nonumber \\
&&\hspace{-0.6cm} + \sum_{i=1}^{+\infty} \! \left[ p_c^{i+\beta}(1-p_c) \!\! \left( \sum_{j=0}^{\beta} \mathbb{E}(\lambda_j) + i \cdot \mathbb{E}(\lambda_{\beta}) \right) \right] \nonumber \\
= &&\hspace{-0.6cm} \sum_{i=0}^{\beta-1} p_c^i \cdot \mathbb{E}(\lambda_i) + \frac{p_c^{\beta}}{1-p_c} \mathbb{E}(\lambda_{\beta}) \nonumber \\
= &&\hspace{-0.6cm} \sum_{i=0}^{\beta-1} p_c^i \frac{1+2^i\text{CW}_\text{min}}{2} + \frac{p_c^{\beta}}{1-p_c} \frac{1+2^{\beta}\text{CW}_\text{min}}{2} .
\end{eqnarray}

It can be shown that user B has the same average cycle time as user A. Therefore, we have $\Psi_\text{CSMA/CA}=\mathbb{E}( \Gamma_A)$, which gives us Theorem \ref{prop:CCT_CSMA_basic}.

\bibliographystyle{IEEEtran}
\bibliography{References}

\begin{thebibliography}{10}
\providecommand{\url}[1]{#1}
\csname url@samestyle\endcsname
\providecommand{\newblock}{\relax}
\providecommand{\bibinfo}[2]{#2}
\providecommand{\BIBentrySTDinterwordspacing}{\spaceskip=0pt\relax}
\providecommand{\BIBentryALTinterwordstretchfactor}{4}
\providecommand{\BIBentryALTinterwordspacing}{\spaceskip=\fontdimen2\font plus
\BIBentryALTinterwordstretchfactor\fontdimen3\font minus \fontdimen4\font\relax}
\providecommand{\BIBforeignlanguage}[2]{{%
\expandafter\ifx\csname l@#1\endcsname\relax
\typeout{** WARNING: IEEEtran.bst: No hyphenation pattern has been}%
\typeout{** loaded for the language `#1'. Using the pattern for}%
\typeout{** the default language instead.}%
\else
\language=\csname l@#1\endcsname
\fi
#2}}
\providecommand{\BIBdecl}{\relax}
\BIBdecl

\bibitem{5461911}
T.~Lan, D.~Kao, M.~Chiang, and A.~Sabharwal, ``An axiomatic theory of fairness in network resource allocation,'' in \emph{IEEE INFOCOM}, 2010.

\bibitem{6517050}
H.~Shi, R.~V. Prasad, E.~Onur, and I.~Niemegeers, ``Fairness in wireless networks: Issues, measures and challenges,'' \emph{IEEE Communications Surveys \& Tutorials}, vol.~16, no.~1, pp. 5--24, 2014.

\bibitem{8665952}
Y.~Yu, T.~Wang, and S.~C. Liew, ``Deep-reinforcement learning multiple access for heterogeneous wireless networks,'' \emph{IEEE Journal on Selected Areas in Communications}, vol.~37, no.~6, pp. 1277--1290, 2019.

\bibitem{koksal2000analysis}
C.~E. Koksal, H.~Kassab, and H.~Balakrishnan, ``An analysis of short-term fairness in wireless media access protocols,'' in \emph{ACM SIGMETRICS}, 2000.

\bibitem{1378897}
G.~Berger-Sabbatel, A.~Duda, O.~Gaudoin, M.~Heusse, and F.~Rousseau, ``Fairness and its impact on delay in 802.11 networks,'' in \emph{IEEE Global Telecommunications Conference}, 2004.

\bibitem{5062017}
M.~Uchida and J.~Kurose, ``An information-theoretic characterization of weighted alpha-proportional fairness,'' in \emph{IEEE INFOCOM}, 2009, pp. 1053--1061.

\bibitem{8902705}
Y.~Yu, S.~C. Liew, and T.~Wang, ``Carrier-sense multiple access for heterogeneous wireless networks using deep reinforcement learning,'' in \emph{IEEE Wireless Communications and Networking Conference Workshop}, 2019.

\bibitem{shao2021federated}
Y.~Shao, D.~G{\"u}nd{\"u}z, and S.~C. Liew, ``Federated edge learning with misaligned over-the-air computation,'' \emph{IEEE Transactions on Wireless Communications}, vol.~21, no.~6, pp. 3951--3964, 2021.

\bibitem{6778811}
Y.~Kim and G.~Hwang, ``Design and analysis of medium access protocol: Throughput and short-term fairness perspective,'' \emph{IEEE/ACM Transactions on Networking}, vol.~23, no.~3, pp. 959--972, 2015.

\bibitem{5062022}
M.~Bredel and M.~Fidler, ``Understanding fairness and its impact on quality of service in \textsc{IEEE} 802.11,'' in \emph{IEEE INFOCOM}, 2009.

\bibitem{6816520}
C.~Guo, M.~Sheng, X.~Wang, and Y.~Zhang, ``Throughput maximization with short-term and long-term \textsc{J}ain's index constraints in downlink \textsc{OFDMA} systems,'' \emph{IEEE Transactions on Communications}, vol.~62, no.~5, pp. 1503--1517, 2014.

\bibitem{shao2020flexible}
Y.~Shao and S.~C. Liew, ``Flexible subcarrier allocation for interleaved frequency division multiple access,'' \emph{IEEE Transactions on Wireless Communications}, vol.~19, no.~11, pp. 7139--7152, 2020.

\bibitem{berger2005short}
G.~Berger-Sabbatel, A.~Duda, M.~Heusse, and F.~Rousseau, ``Short-term fairness of 802.11 networks with several hosts,'' in \emph{Mobile and Wireless Communication Networks}, 2005.

\bibitem{shao2022learning}
Y.~Shao, Y.~Cai, T.~Wang, Z.~Guo, P.~Liu, J.~Luo, and D.~Gunduz, ``Learning-based autonomous channel access in the presence of hidden terminals,'' \emph{IEEE Transactions on Mobile Computing}, 2023.

\bibitem{jain1984quantitative}
R.~K. Jain, D.-M.~W. Chiu, W.~R. Hawe \emph{et~al.}, ``A quantitative measure of fairness and discrimination,'' \emph{Eastern Research Laboratory, Digital Equipment Corporation, Hudson, MA}, vol.~21, 1984.

\bibitem{8746330}
M.~M. Al-Wani, A.~Sali, B.~M. Ali, A.~A. Salah, K.~Navaie, C.~Y. Leow, N.~K. Noordin, and S.~J. Hashim, ``On short term fairness and throughput of user clustering for downlink non-orthogonal multiple access system,'' in \emph{IEEE Vehicular Technology Conference}, 2019.

\bibitem{4346554}
B.~Radunovic and J.-Y. Le~Boudec, ``A unified framework for max-min and min-max fairness with applications,'' \emph{IEEE/ACM Transactions on Networking}, vol.~15, no.~5, pp. 1073--1083, 2007.

\bibitem{shao2020significant}
Y.~Shao, A.~Rezaee, S.~C. Liew, and V.~W. Chan, ``Significant sampling for shortest path routing: A deep reinforcement learning solution,'' \emph{IEEE Journal on Selected Areas in Communications}, vol.~38, no.~10, pp. 2234--2248, 2020.

\bibitem{kelly1997charging}
F.~Kelly, ``Charging and rate control for elastic traffic,'' \emph{European transactions on Telecommunications}, vol.~8, no.~1, pp. 33--37, 1997.

\bibitem{9356328}
Z.~Jing, Q.~Yang, M.~Qin, J.~Li, and K.~S. Kwak, ``Long-term max-min fairness guarantee mechanism for integrated multi-\textsc{RAT} and \textsc{MEC} networks,'' \emph{IEEE Transactions on Vehicular Technology}, vol.~70, no.~3, pp. 2478--2492, 2021.

\bibitem{li2019adaptive}
Z.~Li, Y.~Bai, J.~Liu, J.~Chen, and Z.~Chang, ``Adaptive proportional fair scheduling with global-fairness,'' \emph{Wireless Networks}, vol.~25, pp. 5011--5025, 2019.

\bibitem{d2020fairness}
A.~D'Amour, H.~Srinivasan, J.~Atwood, P.~Baljekar, D.~Sculley, and Y.~Halpern, ``Fairness is not static: deeper understanding of long term fairness via simulation studies,'' in \emph{Proceedings of the 2020 Conference on Fairness, Accountability, and Transparency}, 2020.

\bibitem{9502043}
IEEE, ``\textsc{IEEE} standard for information technology--telecommunications and information exchange between systems - local and metropolitan area networks--specific requirements - part 11: Wireless \textsc{LAN} medium access control (\textsc{MAC}) and physical layer (\textsc{PHY}) specifications,'' \emph{IEEE Std 802.11-2020}, 2021.

\bibitem{840210}
G.~Bianchi, ``Performance analysis of the \textsc{IEEE} 802.11 distributed coordination function,'' \emph{IEEE Journal on Selected Areas in Communications}, vol.~18, no.~3, pp. 535--547, 2000.

\bibitem{4542784}
O.~Ekici and A.~Yongacoglu, ``\textsc{IEEE} 802.11a throughput performance with hidden nodes,'' \emph{IEEE Communications Letters}, vol.~12, no.~6, pp. 465--467, 2008.

\bibitem{4799015}
T.~Kim and J.-T. Lim, ``Throughput analysis considering coupling effect in \textsc{IEEE} 802.11 networks with hidden stations,'' \emph{IEEE Communications Letters}, vol.~13, no.~3, pp. 175--177, 2009.

\end{thebibliography}

\end{document}